\documentclass[letterpaper]{JHEP3}
\usepackage{amssymb,amsfonts}
\usepackage{cite}
\usepackage{epsfig}

\newcommand{\CC}{{\cal C}}

\newcommand{\CF}{{\cal F}}
\newcommand{\CG}{{\cal G}}
\newcommand{\CH}{{\cal H}}

\newcommand{\CN}{{\cal N}}
\newcommand{\CO}{{\cal O}}
\newcommand{\CP}{{\cal P}}

\newcommand{\CU}{{\cal U}}
\newcommand{\CV}{{\cal V}}

\newcommand{\bk}{{\bf k}}
\newcommand{\bx}{{\bf x}}
\newcommand{\by}{{\bf y}}

\newcommand{\diff}{{\rm Diff}}
\newcommand{\usigma}{U(1)_\Sigma}
\newcommand{\p}{\partial}

\newcommand{\be}{\begin{equation}}
\newcommand{\ee}{\end{equation}}
\newcommand{\bea}{\begin{eqnarray}}
\newcommand{\eea}{\end{eqnarray}}

\newcommand{\ie}{{\it i.e.}}
\newcommand{\eg}{{\it e.g.}}

\usepackage{graphicx}
\usepackage{latexsym}
\title{General Covariance\\ in Quantum Gravity at a Lifshitz Point}
\author{Petr Ho\v{r}ava${}^{a,b,c}$ 
and Charles M. Melby-$\!$Thompson${}^{a,b}$ 
\\ 
${}^a$Berkeley Center for Theoretical Physics and Department of Physics\\
University of California, Berkeley, CA, 94720-7300, USA\medskip\\ 
${}^b$Theoretical Physics Group, Lawrence Berkeley National Laboratory\\
Berkeley, CA 94720-8162, USA\medskip\\ 
${}^c$Institute for the Physics and Mathematics of the Universe\\
University of Tokyo, Kashiwa 277-8568, Japan}
\abstract{In the minimal formulation of gravity with Lifshitz-type anisotropic 
scaling, the gauge symmetries of the system are foliation-preserving 
diffeomorphisms of spacetime.  Consequently, compared to general relativity, 
the spectrum contains an extra scalar graviton polarization.  Here we 
investigate the possibility of extending the gauge group by a local $U(1)$ 
symmetry to ``nonrelativistic general covariance.''  This extended  
gauge symmetry eliminates the scalar graviton, and forces the coupling 
constant $\lambda$ in the kinetic term of the minimal formulation to take its 
relativistic value, $\lambda=1$.  The resulting theory exhibits anisotropic 
scaling at short distances, and reproduces many features of general relativity 
at long distances.}  
\begin{document}
%%%%%%%%%%%%%%%%%%%%%%%%%%%%%%%%%%%%%%%%%%%%%%%%%%%%%%%%%%%%%%%%%%%%%%%%%%%%%%%
\section{Introduction}
\label{seci}

The idea of gravity with anisotropic scaling \cite{mqc,lif,spdim} has 
attracted a lot of attention recently.  There are two, somewhat distinct, 
motivations for developing this approach to gravity.  The first is driven 
by the long-standing search for a theoretical framework in which the classical 
theory of gravity is reconciled with the laws of quantum mechanics.  
A successful outcome of this search would result in a mathematically 
self-consistent framework for quantum gravity, not necessarily subjected to 
experimental tests.  Examples already exist -- the ten-dimensional 
supersymmetric vacua of string theory belong to this category.  The second 
motivation comes from a goal which is more narrow, and also much more 
ambitious:  To find, within such a self-consistent quantum gravity framework, 
a theory that reproduces the observed gravitational phenomena in our universe 
of $3+1$ macroscopic dimensions.  

Both of these motivations are relevant for the development of gravity with 
anisotropic scaling.  For a large class of possible applications, it does not 
matter whether or not the theory matches general relativity at long distances, 
or conforms to the available experimental tests of gravity in $3+1$ 
dimensions.  A mathematically consistent quantum gravity which lacks 
this phenomenological matching can still be useful in the context of AdS/CFT 
correspondence, and produce novel gravity duals for a broader class of 
field theories, of interest for example in condensed matter applications.  
It can also have interesting mathematical implications, given the close 
connection between the theory formulated in \cite{mqc,lif} and the 
mathematical theory of the Ricci flow on Riemannian manifolds.  

However, explaining the observed features of gravity in our universe of $3+1$ 
dimensions is still perhaps the leading motivation for developing a quantum 
theory of gravity.  Therefore, it makes sense to ask how close we can get, 
in the new framework of gravity with anisotropic scaling, to reproducing 
general relativity in the range of scales where the laws of gravity have been 
experimentally tested.  

The comparison to general relativity is facilitated by the fact that 
in the framework proposed in \cite{mqc,lif,spdim}, gravity is also described 
simply as a field theory of the dynamical metric on spacetime.  Unlike in 
general relativity, however, the spacetime manifold $M$ (which we take to 
be of a general dimension $D+1$) is equipped with a preferred structure of a 
codimension-one foliation $\CF$ by slices of constant time, $\Sigma(t)$.%
\footnote{For simplicity, we will assume throughout this paper that the leaves 
$\Sigma(t)$ of this foliation all have the same topology of a $D$-dimensional 
manifold $\Sigma$.}  
In the minimal realization of the theory, reviewed in Section~\ref{secmin}, 
the gauge symmetries of the system are the foliation-preserving 
diffeomorphisms $\diff(M,\CF)$.  Since this symmetry contains one less gauge 
invariance per spacetime point compared to the full spacetime diffeomorphisms 
$\diff(M)$, the spectrum of the linearized theory around flat spacetime 
contains one additional, scalar polarization of the graviton.  

At short distances, the anisotropy between space and time is measured by 
a nontrivial dynamical critical exponent $z>1$, leading to an improved 
ultraviolet behavior of the theory.  At long distances, on the other hand, 
the theory is driven to an infrared regime where it shares many features with 
general relativity.  First of all, under the influence of relevant terms in 
the classical action, the scaling becomes naturally isotropic, with the 
relativistic value of $z=1$.  Moreover, the lowest-dimension terms that 
dominate the action in the infrared are exactly those that appear in the ADM 
decomposition \cite{adm} of the Einstein-Hilbert action:  The scalar curvature 
term, which sets the value of the effective Newton constant, and the 
cosmological constant term.  

Thus, in the low-energy 
%%( limit, 
regime,
%%)
the action of the minimal theory with 
anisotropic scaling looks very similar to that of general relativity.  
However, this similarity has its limits, and the theories are clearly 
different even in the infrared.  The differences can be understood in three 
related ways: As a difference in gauge symmetries, a difference in the 
graviton spectrum, and a difference in the number of independent coupling 
constants.  First, in the theory with anisotropic scaling, the gauge symmetry 
is reduced to $\diff(M,\CF)$, and the theory propagates an extra scalar 
polarization of the graviton.  In addition, the kinetic term in the action 
allows for an additional coupling $\lambda$, which is undetermined by any 
symmetry of the minimal theory, and therefore expected to run with the scale 
in the quantum theory.  In general relativity, the spacetime diffeomorphism 
symmetries force $\lambda=1$ and protect this value from quantum corrections.  
Stringent experimental limits on the value of $\lambda$ have been advocated 
in the literature \cite{niayesh,lambda}, suggesting that at least in this 
class of models, it must be very near its relativistic value $\lambda=1$.  
However, in the regime near $\lambda=1$, difficulties with the dynamics of the 
additional scalar graviton have been pointed out (see, for example, 
\cite{cnps,bpsone,bpstwo}).

In order to get closer to general relativity, it is tempting to focus on the 
structure of gauge symmetries.  However, this needs to be done cautiously, 
keeping in mind that gauge symmetries are just convenient redundancies in the 
description of a physical system, and therefore to some extent in the eye of 
the beholder.  The more physical perspective is to focus on the spectrum of 
propagating degrees of freedom.% 
\footnote{Of course, one can make the theory diffeomorphism invariant 
in a trivial way, by integrating in the gauge invariance without changing 
the number of physical degrees of freedom.  This leads to a theory 
which is formally generally covariant, but effectively equivalent to the 
original model.  Perhaps any sensible theory can be ``parametrized'' in this 
way \cite{ht}, and formally rewritten as a generally covariant theory.  
%%(
This process of covariantizing a given theory is very closely connected to the 
St\"uckelberg mechanism used prominently in particle physics (see, \eg , 
\cite{ruegg} for a review).  It has been applied to the models of gravity with 
anisotropic scaling \cite{mqc,lif} in \cite{cnps,st1,st2}.  In this paper, 
instead, we are interested in the nontrivial extension of gauge symmetry 
which actually reduces the number of physical degrees of freedom.  
Investigating how the resulting theory then responds to the St\"uckelberg 
trick is an interesting question, beyond the scope of this paper.%
%%)
}
Thus, we will be interested in finding an extension of the gauge symmetry that 
will turn the extra, scalar polarization of the graviton into a gauge 
artifact.  A second option would be to find a mechanism for generating a 
finite mass gap for the scalar graviton -- in this paper, we concentrate on 
the first possibility.  

We will find such an extended gauge symmetry, with as many generators 
per spacetime point as in general relativity.  This gauge symmetry can be 
viewed as representing ``nonrelativistic general covariance'' in gravity with 
anisotropic scaling.  The extended symmetry eliminates the scalar polarization 
of the graviton from the spectrum.  As a bonus, we find that the extended 
gauge symmetry requires $\lambda=1$, thereby reducing the kinetic term to 
coincide with that of general relativity.  It is in fact important that our 
entire construction depends only on the form of the kinetic term, and 
therefore does not restrict the form of the potential term in the action.  
Hence, at short distances, the covariant theory can exhibit the same improved 
ultraviolet behavior associated with $z>1$ in the minimal theory of 
\cite{mqc,lif}.  

In this paper, our perspective is that of (effective) quantum field theory, 
but we restrict our analysis to the leading tree-level, or classical, 
approximation.  Quantum corrections are expected to modify the scaling 
behavior of our models, but they are beyond the scope of this paper.  In 
addition, our analysis will be strictly local: We freely integrate by parts 
and ignore total derivative terms.  Boundary terms play a notoriously central 
role in relativistic gravity; it will be important to extend our analysis to 
include their precise structure and clarify their role in theories of gravity 
with anisotropic scaling.  These are among the many interesting issues left 
for future work.

The main result of the paper is the construction of the generally covariant 
gravity with anisotropic scaling, which we present in Section~\ref{secgcl}.  
Sections~\ref{seci}, \ref{secgl} and~\ref{secga} prepare the ground for 
a better understanding of the main results, and explore a few additional 
issues of interest.  
% The reader who prefers to cut to the chase can jump directly to 
% Section~\ref{secgcl}.

\subsection{General covariance}

In order to explain what exactly we mean by ``general covariance,'' we first 
consider two theories -- general relativity, and the ultralocal theory of 
gravity \cite{teitelboim,henneaux} -- and illustrate our point using the 
Hamiltonian formulation of these two theories.  

In fact, throughout this paper we will often resort to the Hamiltonian 
formalism,%
\footnote{For the canonical reference on Hamiltonian systems with constraints, 
see \cite{ht}.} 
for a number of reasons.  First of all, the time versus space split of the 
Hamiltonian formalism is particularly natural for gravity with anisotropic 
scaling.  More importantly, the technology available in the Hamiltonian 
formalism allows us to get a precise count of the number of propagating 
degrees of freedom, and offers a better insight into the structure of the 
gauge symmetries of the theory.  Indeed, one of the advantages of the 
Hamiltonian formulation is that one does not have to specify the gauge 
symmetries {\it a priori}.  Instead, the structure of the Hamiltonian 
constraints provides an essentially algorithmic way in which the correct gauge 
symmetry structure is determined automatically \cite{ht}.  In the process, the 
consistency of the equations of motion is tied to the closure of the 
constraint algebra and the preservation of the constraints under the time 
evolution.  Once the full system of constraints has been 
determined, the constraints are separated into first-class (whose commutators 
with other constraints vanish on the constraint surface) and second-class 
(whose commutators define a nondegenerate symplectic form).    As an 
additional benefit, after determining the numbers $\CC_1$ of first-class and 
$\CC_2$ of second-class constraints, the number of degrees of freedom $\CN$ 
can be reliably evaluated by the standard formula \cite{ht} 
\be
\label{ndof}
\CN=\frac{1}{2}\left(\dim\CP-2\CC_1-\CC_2\right),
\ee
where $\dim\CP$ is the number of fields in the canonical formulation (\ie , 
the dimension of phase space).  In local field theory, this formula can 
be interpreted per spacetime point, giving the number of local degrees of 
freedom.  We will use this formula repeatedly throughout the paper.

In canonical general relativity \cite{dirac1,dirac2,adm} on a 
spacetime manifold $M$ with $D+1$ coordinates $(x^i,t)$, the algebra of 
constraints contains the ``superhamiltonian'' $\CH_\perp(\bx)$ and the 
``supermomentum'' $\CH_i(\bx)$, and the total Hamiltonian is just a sum of 
constraints:
\be
\label{hamgr}
H=\int d^D\bx\left(N\CH_\perp+N^i\CH_i\right).
\ee  
$N$ and $N^i$ are the lapse and shift variables of the metric, and $\CH_\perp$ 
and $\CH_i$ are functions of the spatial components $g_{ij}$ of the metric and 
their canonically conjugate momenta $\pi^{ij}$.  Since the constraints are all 
first-class, they generate gauge symmetries, whose generators are 
\be
\CH(\xi^i)\equiv\int d^D\bx\,\xi^i(\bx,t)\CH_i(\bx,t),\qquad 
\CH_\perp(\xi^0)\equiv\int d^D\bx\,\xi^0(\bx,t)\CH_\perp(\bx,t).
\ee
Their commutation relations are 
well-understood, even though they do not quite yield the naively expected 
spacetime diffeomorphism algebra.  True, the commutator of two $\CH_i$'s 
\be
\label{diracss}
[\CH(\xi^i),\CH(\zeta^j)]=\CH(\xi^k\p_k\zeta^i-\zeta^k\p_k\xi^i)
\ee
reproduces the algebra of spatial diffeomorphisms $\diff(\Sigma)$, and 
\be
\label{diracst}
[\CH(\xi^i),\CH_\perp(\zeta^0)]=\CH_\perp(\xi^k\p_k\zeta^0)
\ee
just states that $\CH_\perp$ transforms correctly under $\diff(\Sigma)$.  
However, the commutator of $\CH_\perp$ with itself gives a field-dependent 
result,
\be
\label{diractt}
[\CH_\perp(\xi^0),\CH_\perp(\zeta^0)]=-\sigma\CH\left(g^{ij}(\xi^0\p_j\zeta^0-
\zeta^0\p_j\xi^0)\right).
\ee
Here $\sigma$ denotes the signature of spacetime: $\sigma=-1$ for general 
relativity in Minkowski signature.  This generalized, Dirac algebra is the 
Hamiltonian manifestation of the original diffeomorphism symmetry (and general 
covariance) of general relativity, with $D+1$ gauge symmetries per spacetime 
point.    

Our second example is the ultralocal theory of gravity, which results from 
dropping the spatial scalar curvature term $R$ in the action of general 
relativity.  Of course, this step selects a preferred foliation $\CF$ of 
spacetime, and therefore violates spacetime diffeomorphism invariance.  
One might naively assume that the symmetry is reduced to the 
foliation-preserving diffeomorphisms $\diff(M,\CF)$.  However, the analysis 
of Hamiltonian constraints reveals a surprising fact 
\cite{teitelboim,henneaux}:  The theory is still gauge invariant under as many 
gauge symmetries per spacetime point as general relativity.  In contrast with 
general relativity, the ultralocal theory exhibits a contracted version of 
the Hamiltonian constraint algebra, with (\ref{diractt}) replaced by its 
$\sigma\to 0$ limit:
\be
\label{thtt}
[\CH_\perp(\xi^0),\CH(\zeta^0)]=0,
\ee
while the remaining commutators (\ref{diracss}) and (\ref{diracst}) stay the 
same. 

The theory is ``generally covariant'' -- it has the same number $D+1$ of 
(nontrivial) local gauge symmetries as general relativity, even though the 
algebra in the $\sigma\to 0$ limit still preserves the preferred spacetime 
foliation structure.  Effectively, the spatial diffeomorphism symmetries have 
been kept intact, but the time reparametrization symmetry has been linearized, 
and its algebra contracted to a local $U(1)$ gauge symmetry.%
\footnote{Throughout this paper, we use the rather loose notation common in 
high-energy physics, and refer to any one-dimensional Abelian symmetry group 
as $U(1)$, regardless of whether or not it is actually compact.  Moreover, we 
use the same notation also for the infinite-dimensional, gauge version of the 
$U(1)$ symmetry.}
Just as the Dirac algebra of Hamiltonian constraints (\ref{diracss}), 
(\ref{diracst}) and (\ref{diractt}) in general relativity is associated with 
the Lagrangian symmetries described by the group of spacetime diffeomorphisms, 
$\diff(M)$, the Teitelboim-Henneaux algebra (\ref{diracss}), (\ref{diracst}) 
and (\ref{thtt}) can be associated with a Lagrangian symmetry group which 
takes the form of a semi-direct product,
\be
\label{nrelgc}
U(1)\ltimes\diff(M,\CF).
\ee
It is natural to interpret (\ref{nrelgc}) as the symmetry group of 
``nonrelativistic general covariance.''  This is the structure of gauge 
symmetries that we will try to implement in the case of gravity with 
anisotropic scaling for general values of $z$.  

The restoration of general covariance characterized by (\ref{nrelgc}) still 
maintains the special status of time, keeping it on a different footing from 
space.  We view the fact that ``time is different'' as a virtue of this 
approach:  Indeed, we are looking for possible concessions on the side of 
general relativity that would make it friendlier to the way in which time is 
treated in quantum mechanics, without changing too much of its elegant 
geometric nature.  

\subsection{The minimal theory of gravity with anisotropic scaling}
\label{secmin}

Here we review the basics of the simplest version of quantum gravity with 
anisotropic scaling \cite{mqc,lif}. 

We assume that the spacetime manifold $M$ is a differentiable manifold 
equipped with the extra structure of a preferred foliation $\CF$ by 
codimension-one leaves $\Sigma$ of constant time.  Our starting point is 
a theory whose only dynamical field is the spatial metric, represented in a 
coordinate system $(\bx\equiv x^i,t)$ on $\Sigma$ by components 
$g_{ij}(\bx,t)$.  In a sense, this is the most ``primitive'' implementation 
of the idea of anisotropic scaling for gravity:  The theory exhibits no 
(time-dependent) gauge symmetries, and consequently the spectrum will contain 
not just the tensor polarizations of the gravitons, but also the vector and 
scalar graviton modes.  

The kinetic term is written using the generalized De~Witt metric%
\footnote{$G^{ijk\ell}$ can be viewed as a metric on the space of symmetric 
2-tensors.  As in \cite{mqc,lif}, we will denote its inverse by 
$\CG_{ijk\ell}$, to distinguish it from $G^{ijk\ell}$ with all four indices 
lowered via $g_{ij}$.}   
\be
\label{dwm}
G^{ijk\ell}=\frac{1}{2}\left(g^{ik}g^{j\ell}+g^{i\ell}g^{jk}\right)
-\lambda g^{ij}g^{k\ell}.
\ee
The parameter $\lambda$ is left undetermined by the symmetries of the theory, 
and plays the role of an additional coupling constant.  

This ``primitive'' theory becomes more interesting when we make it gauge 
invariant under foliation-preserving diffeomorphisms $\diff(M,\CF)$, 
whose generators are
\be
\label{difgen}
\delta t=f(t),\qquad \delta x^i=\xi^i(t,\bx).
\ee
The minimal multiplet of fields now contains $g_{ij}$ plus the lapse function 
$N$ and the shift vector $N_i$, which transform under (\ref{difgen}) as
\bea
\label{trsff}
\delta g_{ij}&=&\p_i\xi^kg_{jk}+\p_j\xi^kg_{ik}+\xi^k\p_kg_{ij},\nonumber\\
\delta N_i&=&\p_i\xi^jN_j+\xi^j\p_jN_i+\dot\xi^jg_{ij}+\dot fN_i+f\dot N_i,\\
\delta N&=&\xi^j\p_jN+\dot fN+f\dot N.\nonumber
\eea
The lapse and shift $N$ and $N_i$, known from the ADM parametrization of 
the spacetime metric in general relativity \cite{adm}, play the role of gauge 
fields of the foliation-preserving diffeomorphism symmetry.  Indeed, 
(\ref{trsff}) shows that $N$ (or, more precisely, $\log N$), and $N^i$ 
transform as gauge fields under the gauge transformations, 
\be
\label{gfdiff}
\frac{\delta N}{N}=\dot f+\ldots,\qquad \delta N^i=\dot\xi^i+\ldots.
\ee
(Here the ``$\ldots$'' stand for the standard Lie-derivative terms in 
(\ref{trsff}).)
As a result, it is natural to assume that $N$ and $N_i$ inherit the same 
dependence on spacetime as the corresponding generators (\ref{difgen}):  
While $N_i(t,\bx)$ is a spacetime field, $N(t)$ is only a function of 
time, constant along the spatial slices $\Sigma$.  Making this assumption 
about the lapse function will lead to the minimal theory of gravity with 
anisotropic scaling.

The covariantization of the ``primitive'' theory is accomplished by replacing 
the spatial volume element with its covariant version, 
\be
\sqrt{g}\to \sqrt{g}\,N,
\ee
and by trading the time derivative of the metric for the extrinsic curvature 
$K_{ij}$ of the leaves of the foliation $\CF$,
\be
\dot g_{ij}\to 2K_{ij}\equiv\frac{1}{N}\left(\dot g_{ij}-\nabla_i N_j
-\nabla_j N_i\right).
\ee

In this fashion, we obtain the minimal realization of the idea of anisotropic 
scaling in gravity.  This minimal theory is sometimes referred to as 
``projectable,'' because the spacetime metric assembled from the ingredients 
$g_{ij}$, $N_i$ and $N$  satisfies the axioms of a ``projectable metric'' on 
$(M,\CF)$, as defined in the geometric theory of foliations.  The action of 
the minimal theory is
\be
\label{smint}
S=\frac{2}{\kappa^2}\int dt\,d^D\bx\,\sqrt{g}\,N\left(K_{ij}K^{ij}-\lambda K^2
-\CV\right),
\ee
where $K=g^{ij}K_{ij}$.  The potential term $\CV$ is an arbitrary 
$\diff(\Sigma)$-invariant local scalar functional built out of the spatial 
metric, its Riemann tensor and the spatial covariant derivatives, without 
the use of time derivatives.  

Around a free-field fixed point with dynamical exponent $z$, 
we will measure the scaling dimensions of fields in the units of the spatial 
momentum: $[\p_i]\equiv 1$.  In these units, the volume element in the action 
is of dimension $[dt\,d^D\bx]=-z-D$, suggesting the natural scaling dimensions 
for the field multiplet, 
\be
\label{scaledim}
[g_{ij}]=0,\qquad [N_i]=z-1,\qquad [N]=0.
\ee
This scaling further implies that $[\kappa^2]=z-D$.

If we wish for the theory to be power-counting renormalizable, it is natural 
to start the analysis of possible terms appearing in $\CV$ at short 
distances.  Around a hypothetical Gaussian fixed point, power-counting 
renormalizability in $D+1$ dimensions requires $\CV$ to be dominated by 
terms with $2D$ spatial derivatives, implying in turn that the dynamical 
critical exponent should be equal to $z=D$.  For example, in $3+1$ dimensions, 
there is a natural potential 
\be
\label{vuv}
\CV_{\rm UV}=w^2C_{ij}C^{ij}+\ldots,
\ee
where $w$ is a dimensionless coupling, and 
$C^{ij}=\varepsilon^{ik\ell}\nabla_k(R^j_\ell-\frac{1}{4}R\delta^j_\ell)$ 
is the Cotton tensor.  In (\ref{vuv}), we have indicated only the part of the 
potential that is dominant in the ultraviolet, with ``$\ldots$'' denoting the 
relevant terms which contain fewer than six spatial derivatives and become 
important at longer distances.  

Under the influence of the relevant 
terms, the theory will flow, until it is dominated at long distances by the 
most relevant terms.  In this regime, it makes sense to reorganize the terms 
in $\CV$ by focusing on those most dominant in the infrared: 
\be
\label{vir}
\CV_{\rm IR}=-\mu^2(R-2\Lambda)+\ldots.
\ee
Here $\mu$ and $\Lambda$ are dimensionful couplings of dimensions $[\mu]=z-1$ 
and $[\Lambda]=2$, and the ``$\ldots$'' now denote all the terms containing 
composite operators of {\it higher\/} dimension compared to the displayed, 
most dominant infrared terms.  

It is useful to note that the algebra of gauge symmetries $\diff(M,\CF)$, 
and their action on the fields, can be obtained simply by taking a 
nonrelativistic reduction of the fully relativistic spacetime diffeomorphism 
symmetry and its action on the relativistic metric $g_{\mu\nu}$.  As we will 
see in Section~\ref{secgeom}, a natural extension of this procedure to 
subleading terms in $1/c$ leads to a natural geometric interpretation of the 
extended symmetries that are the focus of this paper.  

\subsection{Comments on the nonprojectable case}
\label{secnonpj}

The minimal, projectable theory can be rewritten in the Hamiltonian formalism, 
with the Hamiltonian similar to (\ref{hamgr}), 
\be
\label{hamlif}
H=\int d^D\bx\left(N\CH_0+N^i\CH_i\right). 
\ee  
Here $\CH_0$ and $\CH_i$ are again functions of the spatial 
metric and its conjugate momenta.  In fact, $\CH_i$ takes the same form 
as in general relativity, $\CH^i=-2\nabla_k\pi^{ik}$, and $\CH_0$ depends 
on the choice of $\CV$.  The main conceptual difference compared to general 
relativity stems from the fact that because $N(t)$ is independent of the 
spatial coordinates $x^i$, it only gives rise to the integral constraint 
$\int d^D\bx\,\CH_0=0$.  Consequently, compared to general relativity, the 
number of first-class constraints and hence gauge symmetries per spacetime 
point has been reduced by one.  

The first, most naive temptation how to eliminate this discrepancy and get 
closer to general relativity is to restore the full dependence of the lapse 
function on space and time by hand.  This option, often referred to in the 
literature as the ``nonprojectable case'' \cite{mqc,lif}, can be viewed at 
least from two different perspectives, which lead to different results.  

First, one can follow the logic of effective field theory:  Having postulated 
a multiplet of spacetime fields 
\be
g_{ij}(t,\bx),\qquad N_i(t,\bx),\qquad N(t,\bx),
\ee
we postulate a list of global and gauge symmetries, and construct the most 
general action allowed.  In the case at hand, the natural gauge symmetries are 
the foliation-preserving diffeomorphisms $\diff(M,\CF)$.  
While (\ref{hamlif}) is invariant under $\diff(M,\CF)$, {\it it is not the 
most general Hamiltonian compatible with these gauge symmetries.}  As was 
pointed out already in \cite{mqc,lif} and further elaborated in 
\cite{bpsone,bpstwo}, in this effective field theory approach 
to the nonprojectable theory, all terms compatible with the gauge symmetry 
should be allowed in the Lagrangian.  Promoting the lapse function to a 
spacetime field gives a new ingredient for constructing gauge-invariant terms 
in the action,
\be
\nabla_iN/N,
\ee
which transforms under $\diff(M,\CF)$ as a spatial vector and a time scalar.  
Once terms with this new ingredient are allowed in the action, the Hamiltonian 
is no longer linear in $N$, but the algebra of constraints is well-behaved 
\cite{jozka}.  The constraint implied by the variation of 
$N$ is now second-class, and the expected number of propagating degrees of 
freedom is the same as in the minimal theory.

Another possible interpretation of the nonprojectable theory also starts 
by promoting $N$ to a spacetime-dependent field $N(t,\bx)$.  Instead of 
specifying {\it a priori\/} gauge symmetries, however, one can postulate that 
the Hamiltonian take the form (\ref{hamlif}), linear in $N$ \cite{henx}.  
This step must be followed by the analysis of the algebra of Hamiltonian 
constraints, which determines {\it a posteriori\/} whether this construction 
is consistent, and if so, what is the resulting structure of the gauge 
symmetries.  Here the difficulty is in closing the constraint algebra 
\cite{mqc,rome,mli,henx} (see also \cite{emil}):  For general $\CV$, the 
commutator of $\CH_0(\bx)$ with $\CH_0(\by)$ is a complicated function of 
all variables, and the requirement of closure is difficult to implement.  
One interesting exception has been found in the infrared limit \cite{rest} 
(see also \cite{henx,% 
%%(
pons%
%%)
}):  Adding $\pi\equiv g_{ij}\pi^{ij}$ as another 
constraint closes the algebra, turning $\pi$ and $\CH_0$ into a pair of 
second-class constraints.  This infrared theory can then be interpreted as 
general relativity whose gauge freedom has been partially fixed.  This is a 
very appealing picture, but the problem is that it cannot be straightforwardly 
extended to the full theory beyond the infrared limit.  However, as we will 
see below, the possibility of interpreting $\pi$ as an additional constraint 
in the infrared regime as suggested in \cite{rest} will be echoed in the 
generally covariant theory which we present in Section~\ref{secgcl}.

In addition to the two perspectives just reviewed, there is another option how 
to close the constraint algebra in the nonprojectable theory, and interpret it 
as a topological field theory.%
\footnote{This option was pointed out by one of us in \cite{rome}; see also 
Section~5.4 of \cite{mqc}.}
This is possible when the theory satisfies the detailed balance condition 
\cite{mqc,lif}, \ie , when the potential $\CV$ in (\ref{smint}) is of the 
special form 
\be
\CV=\frac{1}{4}\CG_{ijk\ell}\frac{\delta W}{\delta g_{ij}}
\frac{\delta W}{\delta g_{k\ell}}
\ee
for some action functional $W(g_{ij})$ which depends only on $g_{ij}$ and 
its spatial derivatives.  In such cases, it is convenient to introduce a 
system of complex variables, defined as
\be
a^{ij}={\rm i}\pi^{ij}+\frac{1}{\kappa^2}\frac{\delta W}{\delta g_{ij}},\qquad 
\bar a{}^{ij}=-{\rm i}\pi^{ij}+\frac{1}{\kappa^2}\frac{\delta W}{\delta 
g_{ij}}.
\ee
Under the Poisson bracket, these variables play essentially the role of 
a creation and annihilation pair, their only nonzero bracket being
\be
\label{aacomm}
[a^{ij}(\bx),\bar a{}^{k\ell}(\by)]=-\frac{2{\rm i}}{\kappa^2}
\frac{\delta^2 W}{\delta g_{ij}(\bx)\delta g_{k\ell}(\by)}.
\ee  

The Hamiltonian constraints $\CH_i$ and $\CH_0$ can be expressed as simple 
functions of the complex variables,
\be
\CH^i={\rm i}\nabla_j\left(a^{ij}-\bar a{}^{ij}\right),\qquad 
\CH_0=\frac{\kappa^2}{2}a^{ij}\CG_{ijk\ell}\bar a{}^{k\ell}.
\ee
The problematic commutator of $\CH_0(\bx)$ and $\CH_0(\by)$ is still rather  
complicated, but it clearly vanishes when $a^{ij}$ or $\bar a{}^{ij}$ vanish.  
The constraint algebra can thus be closed by declaring $\CH_i$, together with 
either $a^{ij}$ or $\bar a{}^{ij}$ to be the primary constraints.  This would  
then guarantee that the original Hamiltonian constraints $\CH_i$ and $\CH_0$, 
as well as all their commutators, are zero on the constraint surface.   

This step can be made more precise as follows.  Because $a^{ij}$ and $\bar 
a{}^{ij}$ are complex conjugates of each other, it is not possible to declare 
only (say) $a^{ij}$ to be first-class constraints, at least not without 
making $a^{ij}$ and $\bar a{}^{ij}$ formally independent.  Instead, we 
accomplish our goal by declaring both $a^{ij}$ and $\bar a{}^{ij}$ as 
constraints.  Because their commutator (\ref{aacomm}) is nonzero, these 
constraints are second-class and do not imply any additional gauge symmetry.  

However, a pair of second-class constraints can often be interpreted as 
a first-class constraint, together with a gauge-fixing condition.  We can 
interpret the theory with the second-class constraints $a^{ij}$ and 
$\bar a{}^{ij}$ in this fashion:  First, we choose $a^{ij}-\bar a{}^{ij}$ 
as the first-class constraint.  The gauge symmetry generated by this 
constraint acts on $g_{ij}$ via 
\be
\label{topos}
\delta g_{ij}=\lambda_{ij}(t,\bx),
\ee
with $\lambda_{ij}$ an arbitrary spacetime-dependent symmetric two-tensor.  
This is just the topological gauge symmetry as introduced originally by Witten 
\cite{wtft,wcft}, here acting on the spatial component of the metric.  The 
theory is then fully specified by the choice of a gauge-fixing condition for 
the topological gauge symmetry.  In our case, this choice should restore 
$a^{ij}$ and $\bar a{}^{ij}$ as second-class constraints.  Choosing 
$a^{ij}+\bar a{}^{ij}$ as the gauge fixing condition is certainly a consistent 
possibility; however, a more interesting scenario is available when we 
Wick-rotate the theory to imaginary time, $t=-i\tau$.  This case is of 
particular interest because topological field theories are typically formulated 
in imaginary time.  In this regime, $a^{ij}$ and $\bar a{}^{ij}$ are now real, 
instead of being complex conjugates.  We can then select an asymmetric 
gauge-fixing condition, for example
\be
a^{ij}\equiv -\pi^{ij}+\frac{1}{\kappa^2}\frac{\delta W}{\delta g_{ij}}=0.
\ee
This equation is a flow equation for $g_{ij}$ as a function of the imaginary 
time $\tau$, reminiscent of the Ricci flow equation and its cousins.  We have 
thus obtained a topological field theory associated with the flow equations 
on Riemannian manifolds.  Indeed, the number of topological gauge symmetries 
(\ref{topos}) is the same as the number of field components of $g_{ij}$: 
The theory has no local propagating degrees of freedom in the bulk.  

The original $\diff(M,\CF)$ symmetry can be viewed as a separate gauge 
symmetry in addition to the topological gauge symmetry (\ref{topos}).  
However, because the action (\ref{trsff}) of $\diff(M,\CF)$ on $g_{ij}$ is a 
special case of (\ref{topos}), including $\diff(M,\CF)$ explicitly leads to a 
redundancy in gauge symmetries, and triggers the appearance of 
``ghost-for-ghosts'' in the BRST formalism.  In this respect, the structure 
of the gauge symmetries is very similar to the conventional topological field 
theories of the cohomological type \cite{wtft,wcft} such as topological 
Yang-Mills theory.  

In this paper, we are interested in gravity with bulk propagating degrees of 
freedom, whose spectrum contains the tensor polarizations of the graviton but 
not the scalar mode.  Therefore, we do not pursue the nonprojectable theory 
further, and look for the missing gauge invariance elsewhere.

\section{Global $\usigma$ Symmetry in the Minimal Theory at $\lambda=1$}
\label{secgl}

In general relativity, the value $\lambda=1$ of the coupling in (\ref{smint}) 
is selected -- and protected from renormalization -- by the gauge symmetries 
$\diff(M)$ of the theory.  It is perhaps surprising that the case of 
$\lambda=1$ plays a special role in the minimal theory with anisotropic 
scaling as well \cite{mqc,lif}.  One can see this by examining the 
spectrum of the linarized fluctuations around the flat space solution.  

For simplicity, we will now assume that the flat spacetime
\be
\label{flatst}
g_{ij}=\delta_{ij},\qquad N_i=0,\qquad N=1
\ee
is a solution of the equations of motion, and expand the metric to linear 
order around this background,
\be
g_{ij}=\delta_{ij}+\kappa h_{ij},\qquad N_i=\kappa n_i,\qquad N=1+\kappa n.
\ee
Since $n$ is not a spacetime field but only a function of time, its equation 
of motion gives one integral constraint, and does not affect the number of 
local degrees of freedom or their dispersion relations.  Therefore, we 
only consider the equations of motion for the spacetime fields $h_{ij}$ and 
$n_i$. 
 
It will be convenient to further decompose the $h_{ij}$ and $n_i$ 
fluctuations into their irreducible components, 
\be
\label{irrflg}
h_{ij}=s_{ij}+\p_iw_j+\p_jw_i
+\left(\p_i\p_j-\frac{1}{D}\delta_{ij}\p^2\right)B+\frac{1}{D}\delta_{ij}h,
\ee
where the scalar $h=h_{ii}$ is the trace part of $h_{ij}$, while $s_{ij}$ is 
symmetric, traceless and transverse (\ie , divergence-free: $\p_i s_{ij}=0$), 
and $w_i$ is transverse; and similarly, 
\be
\label{irrfln}
n_i=u_i+\p_iC,
\ee
with $u_i$ transverse, $\p_iu_i=0$.  
It is also useful to decompose the linearized gauge transformations,
\be
\xi^i(\bx,t)=\zeta_i(\bx,t)+\p_i\eta(\bx,t).  
\ee
In this decomposition, $\zeta^i$ satisfy $\p_i\zeta_i=0$, and therefore 
represent the generators of linearized volume-preserving spatial 
diffeomorphisms.  The linearized gauge transformations act on the irreducible 
components of the fields via
\bea
\label{lingt}
\delta s_{ij}=0,\qquad \delta w_i&=&\zeta_i,\qquad\delta B=2\eta,\qquad\delta
h=2\p^2\eta,\cr
\delta u_i&=&\dot\zeta_i,\qquad\delta C=\dot\eta.
\eea
These rules suggest a few natural gauge-fixing conditions.  For example, 
we can set $u_i=0$  and $C=0$, which leaves the residual symmetries with 
time-independent $\zeta_i(\bx)$ and $\eta(\bx)$, or $w_i=0$ and $B=0$, which 
fixes the gauge symmetries completely.  In either gauge, the spectrum of 
linearized fluctuations around the flat background contains transverse 
traceless polarizations $s_{ij}$ which all share the same dispersion 
relation (dependent on the details of $\CV$), and a scalar whose dispersion 
is dependent on $\lambda$.  In the vicinity of $\lambda=1$, the dispersion 
relation of the scalar graviton exhibits a singular behavior,
\be
\label{dispone}
\omega^2=(\lambda-1)F(\bk^2,\lambda),
\ee
where $F(\bk^2,\lambda)$ is a regular function of $\lambda$ near $\lambda=1$, 
whose details again depend on $\CV$.  Thus, the scalar dispersion relation 
degenerates to $\omega^2=0$ in the limit of $\lambda\to 1$ \cite{mqc,lif}.  

\subsection{Symmetries in the linearized approximation around flat spacetime}
\label{secsy}

The spectrum of linear excitations around the flat spacetime shows that 
the relativistic value $\lambda=1$ is special even in the nonrelativistic 
theory, as indicated by the dispersion relation of the scalar graviton mode 
(\ref{dispone}) which degenerates as $\lambda\to 1$.  This singular behavior 
was explained in \cite{mqc}:  At $\lambda=1$, the linearized theory with 
$\lambda=1$ enjoys an interesting Abelian symmetry, which acts on the fields 
of the minimal theory via
\be
\label{symrig}
\delta n_i=\p_i\alpha,\qquad \delta h_{ij}=0,\qquad \delta n=0.
\ee
Here the parameter $\alpha(\bx)$ is an arbitrary smooth function of the 
spatial coordinates, constant in time: 
\be
\label{alphadot}
\dot\alpha=0.
\ee
Since the generator $\alpha$ is independent of time, it is natural to 
interpret this infinite-dimensional Abelian symmetry as a {\it global\/} 
symmetry:  In the nonrelativistic setting, it is the hallmark of gauge 
symmetries in the Lagrangian formalism that their generators are arbitrary 
functions of time.  In order to indicate that the Abelian symmetries generated 
by $\alpha(\bx)$ represent a collection of $U(1)$ symmetries parametrized 
by the spatial slice $\Sigma$ of the spacetime foliation, we will refer to 
this infinite-dimensional symmetry by $\usigma$.

At this stage, it is interesting to note that the $\usigma$ symmetry looks 
very reminiscent of a residual gauge symmetry 
in a gauge theory, in which some sort of temporal gauge has been chosen.  
As we will see in the rest of the paper, this intuition is essentially 
correct, but the specific realization of this idea in the full nonlinear 
theory will be surprisingly subtle.

In order to see that $\usigma$ is indeed a symmetry of the linearized theory 
at $\lambda=1$, it is instructive to restore temporarily $\lambda$, and 
evaluate the variation of the action under (\ref{symrig}) in the linearized 
approximation (which we denote by ``$\approx$''), while allowing 
$\alpha$ to be time dependent:
\be
\label{varqa}
\delta_\alpha S\approx 2\int dt\,d^D\bx\left\{\dot\alpha\left(\frac{D-1}{D}
(\p^2)^2 B+\frac{1-\lambda D}{D}\p^2 h\right)-2\alpha (\lambda-1)(\p^2)^2C
\right\}.
\ee
At $\lambda=1$, the last term drops out, and the action is invariant under 
time-independent $\alpha$.  Note also that the term proportional to 
$\dot\alpha$ in (\ref{varqa}) is not gauge invariant under (\ref{lingt}),  
unless $\lambda=1$ when it equals 
\be
\frac{D-1}{D}\left\{(\p^2)^2 B-\p^2 h\right\}\approx R,
\ee
which we recognize as the linearized Ricci scalar of $g_{ij}$.  

Given this global $\usigma$ symmetry, it is natural to ask whether it can 
be gauged.  At $\lambda=1$, this process can be easily completed in the 
linearized theory.  We promote $\alpha$ to an arbitrary smooth function 
of $\bx$ and $t$, introduce a gauge field $A(\bx,t)$, and postulate its 
transformation rules under the gauge transformations,
\be
\delta_\alpha A=\dot\alpha.
\ee
The gauging is accomplished by augmenting the action by a coupling of $A$ to 
the linearized Ricci scalar, 
\be
S_A=-\frac{2(D-1)}{D}\int dt\,d^D\bx\,A\,\left\{(\p^2)^2 B-\p^2 h\right\}.
\ee
It is easy to see that in the gauged theory, the scalar mode of the graviton 
has been eliminated from the spectrum of physical excitations:  With 
$A=0$ as our gauge choice, the equations of motion are the same as in the 
original theory with the global $\usigma$ symmetry, plus the Gauss constraint 
\be
(\p^2)^2B-\p^2h=0.  
\ee
This Gauss constraint eliminates the scalar degree of freedom, leaving only 
the tensor modes of the graviton in the physical spectrum of the linearized 
theory.

\subsection{The nonlinear theory}
\label{secnonl}

We would now like to extend the success of the $U(1)$ gauging from the 
linearized approximation to the full nonlinear theory.  Before we can 
proceed with the gauging, however, we must first check whether $\usigma$ 
extends to a global symmetry of the nonlinear theory.  

In the linearized theory before gauging, the parameter $\alpha(\bx)$ of the 
infinitesimal $\usigma$ transformation was independent of time, and 
consequently we interpreted $\usigma$ as a global symmetry.  In the nonlinear 
theory, the linearized transformation of $n_i$ in (\ref{symrig}) simply 
becomes%
\footnote{The explicit multiplicative factor of $N$ in (\ref{delni}) is 
explained by the requirement of matching the tensorial properties of both 
sides in (\ref{delni}) under $\diff(M,\CF)$.  Thus, it is in fact 
$A_i\equiv N_i/N$ that transforms as the spatial projection of a spacetime 
vector field under $\diff(M,\CF)$ and as the spatial part of a gauge potential 
under $U(1)$, with $\delta A_i=\nabla_i\alpha$.}
\be
\label{delni}
\delta_\alpha N_i=N\nabla_i\alpha.  
\ee
However, the condition (\ref{alphadot}) expressing the time independence of 
$\alpha$ is not covariant under $\diff(M,\CF)$.  The correct covariant 
generalization takes the following modified form,
\be
\label{atimeind}
\dot\alpha-N^i\nabla_i\alpha=0.
\ee
This condition of vanishing covariant time derivative of $\alpha$ is indeed 
invariant under $\diff(M,\CF)$.  

In the full nonlinear theory, the gauge field will transform as a spatial 
scalar and a time vector under $\diff(M,\CF)$, 
\be
\label{gtanl}
\delta A=\dot f A+f\dot A+\xi^i\p_iA,  
\ee
and the gauge transformation of the gauge field becomes
\be
\label{gtfa}
\delta_\alpha A=\dot\alpha-N^i\nabla_i\alpha.
\ee
It follows from (\ref{gtfa}) and (\ref{scaledim}) that the scaling dimensions 
of $\alpha$ and $A$ are given by
\be
\label{scaledima}
[\alpha]=z-2,\qquad [A]=2z-2.
\ee

In the process of evaluating the variation of the action under the general 
$\alpha$ transformation, we will encounter a particular combination of the 
second spatial derivatives of $\dot g_{ij}$, which can be expressed as the 
trace of the time derivative of the Ricci tensor:
\be
\label{rddid}
g^{ij}\dot R_{ij}=\left(g^{ik}g^{j\ell}-g^{ij}g^{k\ell}\right)\nabla_i\nabla_j
\dot g_{k\ell}.
\ee
This formula also implies 
\be
\label{rdotid}
(\sqrt{g}R)\dot{}=-\sqrt{g}\left(R^{ij}-\frac{1}{2}Rg^{ij}\right)\dot g_{ij}+
\sqrt{g}\left(g^{ik}g^{j\ell}-g^{ij}g^{k\ell}\right)\nabla_i\nabla_j
\dot g_{k\ell}.
\ee

It is now straightforward to see that there is an obstruction against 
extending the global symmetry to the full nonlinear theory, at least in 
dimensions greater than $2+1$.  Indeed, for the variation of the action we 
get 
% (setting $\kappa=1$ for simplicity)
%
\bea
\label{varone}
\delta_\alpha S&=&-\frac{1}{\kappa^2}\int dt\,d^D\bx\,\sqrt{g}\left(\dot g_{ij}
-\nabla_iN_j-\nabla_jN_i\right)\left(g^{ik}g^{j\ell}-g^{ij}g^{k\ell}\right)\left(
\nabla_k\nabla_\ell\alpha+\nabla_\ell\nabla_k\alpha\right)\nonumber\\
&&\qquad{}=-\frac{2}{\kappa^2}\int dt\,d^D\bx\,\sqrt{g}\,\alpha\left(g^{ij}
\dot R_{ij}-2G^{ijk\ell}\nabla_k\nabla_\ell\nabla_iN_j\right),
\eea
where in the second line we have integrated by parts twice, dropped the 
corresponding spatial derivative terms, and used (\ref{rddid}).  
The last, triple-derivative term in (\ref{varone}) can be simplified using 
$$G^{ijk\ell}\nabla_k\nabla_\ell\nabla_iN_j=-\nabla^j[\nabla_j,\nabla_k]N^k+
\frac{1}{2}[\nabla_k,\nabla_j]\nabla^jN^k=\nabla_j(R^{jk}N_k),$$
which yields 
\be
\delta_\alpha S=-\frac{2}{\kappa^2}\int dt\,d^D\bx\,\sqrt{g}\,\alpha\left\{
g^{ij}\dot R_{ij}-2\nabla_j(R^{jk}N_k)\right\}.
\ee
Finally, after using the contracted Bianchi identity in the second term, 
integrating by parts in both terms, using (\ref{rdotid}) and dropping the 
total derivatives, we obtain
\bea
\label{varfin}
\delta_\alpha S&=&\frac{2}{\kappa^2}\int dt\,d^D\bx\,\sqrt{g}\left(\dot\alpha
-N^i\nabla_i\alpha\right)R\nonumber\\
&&\qquad\qquad{}-\frac{2}{\kappa^2}\int dt\,d^D\bx\,\sqrt{g}\,\alpha\left(R^{ij}
-\frac{1}{2}Rg^{ij}\right)\left(\dot g_{ij}-\nabla_iN_j-\nabla_jN_i\right).
\eea
The first line in (\ref{varfin}) vanishes for the covariantly time-independent 
$\alpha$ by virtue of (\ref{atimeind}), but the second line represents an 
obstruction against the invariance of $S$, even when $\alpha$ is restricted 
to be covariantly time-independent.  

Another way of seeing the origin of the nonlinear obstruction against the 
$\usigma$ invariance of the minimal theory is the following.  On the 
components $N_i$ of the shift vector, the $\usigma$ transformations act as 
gauge transformations on the components of an Abelian connection.  Define
\be
F_{ij}=\p_iN_j-\p_jN_i.
\ee
Clearly, this is the field strength of $N_i$ interpreted as a connection 
associated with $\usigma$.  Thus, $F_{ij}$ are invariant under $\usigma$, and 
transform as components of a two-form under $\diff(\Sigma)$.  However, 
$F_{ij}$ do {\it not}\/ transform as two-form components under time-dependent 
spatial diffeomorphisms.  

In this new notation, the action with $\lambda=1$ can be rewritten as
\bea
\label{actff}
S&=&\frac{1}{2\kappa^2}\int dt\,d^D\bx\,\frac{\sqrt{g}}{N}\left\{\dot g_{ij}
\left(g^{ik}g^{j\ell}-g^{ij}g^{k\ell}\right)\dot g_{k\ell}-F^{ij}F_{ij}\right.
\nonumber\\
&&\qquad\left.{}-4 R^{ij}N_iN_j+4N^i\left(\nabla^j\dot g_{ij}
-g^{jk}\nabla_i\dot g_{jk}\right)\right\}-\frac{2}{\kappa^2}\int dt\,d^D\bx\,
\sqrt{g}\,N\CV.
\eea
While the first two terms in (\ref{actff}) and the potential term are 
manifestly invariant under $U(1)_\Sigma$, the terms with explicit factors of 
$N_i$ -- which are required by the requirement of $\diff(M,\CF)$ invariance 
-- are not, and their variation reproduces (\ref{varfin}).   Intuitively, 
the obstruction can be related to the fact that $N_i$ plays a dual role in the 
theory.  First, as we have seen in (\ref{gfdiff}),  $N_i$ is the gauge field of 
the time-dependent spatial diffeomorphisms along $\Sigma$.  The second role is 
asked of $N_i/N$ in our attempt to extend the gauge symmetry, and make $N_i/N$ 
transform as the spatial components of a $U(1)$ gauge field.

\section{Gauging the $\usigma$ Symmetry: First Examples}
\label{secga}

Our intention is to gauge the action of the global $\usigma$ in the minimal 
theory with $\lambda=1$.  As we have seen in Section~\ref{seclin}, such 
gauging is possible in the linearized approximation, and it has the desired 
effect of eliminating the scalar polarization of the graviton.  However, in 
Section~\ref{secnonl} we found an obstruction which prevents $\usigma$ from 
being a global symmetry of the minimal theory at the nonlinear level, and 
therefore precludes its straightforward gauging.  More precisely, we found 
that the variation (\ref{varfin}) of the action under an infinitesimal $U(1)$ 
gauge transformation $\alpha(t,\bx)$ consists of two parts,
\be
\label{pone}
\frac{2}{\kappa^2}\int dt\,d^D\bx\,\sqrt{g}\left(\dot\alpha-N^i\nabla_i\alpha
\right)R
\ee
and
\be
\label{ptwo}
-\frac{2}{\kappa^2}\int dt\,d^D\bx\,\sqrt{g}\,\alpha\left(R^{ij}
-\frac{1}{2}Rg^{ij}\right)\left(\dot g_{ij}-\nabla_iN_j-\nabla_jN_i\right).
\ee
If $\usigma$ were a global symmetry, the first step of the Noether procedure 
would be to add a Noether coupling term to the action,
\be
\label{nth}
S_A=-\frac{2}{\kappa^2}\int dt\,d^D\bx\,\sqrt{g}\,AR.
\ee
The $U(1)$ variation (\ref{gtfa}) of $A$ in (\ref{nth}) indeed cancels 
(\ref{pone}).  However, at this stage the Noether procedure breaks down, and 
(\ref{ptwo}) represents the obstruction against gauge invariance. 

In order to allow a straightforward gauging on $\usigma$, we will have to 
find a mechanism which eliminates this obstruction.  Before proceeding with 
that, we first examine the geometric origin of $\usigma$ as a natural gauge 
symmetry, and consider several illustrative cases in which the gauging can 
be completed because the obstruction automatically vanishes.  These include 
a generally covariant nonrelativistic gravity theory in $2+1$ dimensions, and 
an interacting Abelian theory of gravity in general dimensions.  

\subsection{Geometric interpretation of the $U(1)$ symmetry}
\label{secgeom}

The transformation rules (\ref{trsff}) of $\diff(M,\CF)$ on the gravity fields 
can be systematically derived \cite{mqc,lif} from the action of relativistic 
diffeomorphisms $\diff(M)$ on the spacetime metric $g_{\mu\nu}$, as the leading 
order in the nonrelativistic $1/c$ expansion. 

It was already noted in \cite{lif} that the gauge field $A(t,\bx)$ and the 
$U(1)$ symmetry -- which we introduced in a rather {\it ad hoc\/} fashion in 
Sections~\ref{secgl} and~\ref{secga} -- both acquire a natural geometric 
interpretation in the framework of the $1/c$ expansion:  It turns out that 
$A$ is simply the subleading term in the $1/c$ expansion of the relativistic 
lapse function, and $U(1)$ corresponds to the subleading, linearized 
spacetime-dependent time reparametrization symmetry of the relativistic 
theory.  In this section, now make these observations more precise.  

In Section~\ref{secnonpj}, we reviewed some of the difficulties faced in the 
attempts to promote the lapse function to a spacetime field, 
\be
\label{nn}
N(t)\to N(t,\bx).
\ee
In physical terms, the attempts to restore $N$ as a spacetime field can be 
motivated by the desire to restore the information carried by the Newton 
potential in general relativity (for generic gauge choices).  The geometric 
understanding of the gauge field $A$ and the gauge symmetry $U(1)$ shows how 
the generally covariant theory with $U(1)\ltimes\diff(M,\CF)$ symmetry 
restores the Newton potential and avoid the difficulties of the nonprojectable 
theory:  Instead of promoting $N(t)$ into a spacetime field as in (\ref{nn}), 
we keep $N(t)$ as the leading term of the lapse, and introduce the 
{\it subleading\/} term $A(t,\bx)$ in the $1/c$ expansion, at the order in 
which the Newton potential enters in the nonrelativistic approximation to 
general relativity:
\be
\label{nna}
N(t)\to N(t)-\frac{1}{c^2}A(t,\bx).
\ee
Thus, it is only the subleading part of lapse that becomes a spacetime field.  

It is useful to stress that the $1/c$ formalism of this section is just a 
trick, whose sole purpose is to provide a geometric explanation of the action 
of $U(1)\ltimes\diff(M,\CF)$, by taking the formal $c\to\infty$ limit of the 
relativistic $\diff(M)$ symmetry.  The ``speed of light'' $c$ is a formal 
expansion parameter, and should not be confused with the physical speed of 
light which will be generated in our theory at large distances as a result of 
the relevant deformations.   

\subsubsection{The gauge field $A$ and the Newton potential}

In order to reproduce the field content of the theory, and the transformation 
rules under the gauge symmetries, we start with a relativistic spacetime 
metric, and expand it in the powers of $1/c$ as follows:
\be
\label{metrel}
g_{\mu\nu}=\pmatrix{\displaystyle{-N^2 +\frac{N_iN^i}{c^2}+\frac{2NA}{c^2}
+\ldots,}& \displaystyle{\frac{N_i}{c}+\ldots}\cr & \cr
\displaystyle{\frac{N_i}{c} +\ldots,} &  g_{ij}+\ldots}
\ee
This step is complemented by a similar expansion of the relativistic 
spacetime diffeomorphisms with generators $\zeta^\mu$,
\be
\zeta^0=cf(\bx,t)-\frac{1}{c}\frac{\alpha(\bx,t)}{N}+\ldots,\qquad
\zeta^i=\xi^i(\bx,t)+\ldots.  
\ee
In both cases, ``$\ldots$'' refer to terms suppressed by $1/c^2$ compared to 
those displayed.  In the transformation rules, the derivative with respect 
to the relativistic time coordinate is written as $\p/\p x^0=(1/c)\p/\p t$;  
it is then the nonrelativistic time $t$ which is held fixed as $c\to\infty$.  

Taking the $c\to\infty$ limit first requires $\p_if=0$, which means that the 
infinitesimal time reparametrizations $f(t)$  are restricted to be only 
functions of time.  In addition, in accord with (\ref{nna}), we insist that 
$N$ be only a function of time.  The transformation rules (\ref{trsff}) under 
the foliation-preserving diffeomorphisms then follow from the $c\to\infty$ 
limit of the spacetime diffeomorphism symmetry.  In addition, we get
\bea
\delta_\alpha\left(\frac{N_i}{N}\right)&=&\nabla_i\alpha,\cr
\delta_\alpha A&=&\dot\alpha-N^i\nabla_i\alpha,
\eea
with all other fields invariant under $\delta_\alpha$.  We see that the $U(1)$ 
gauge symmetry of interest is geometrically 
interpreted as the subleading part of time reparametrizations in the 
nonrelativistic limit of spacetime diffeomorphisms in general relativity.  

This embedding of the gauge field $A$ into the geometric framework of the 
$1/c$ expansion sheds additional light on the physical role of $A$ in the 
theory.  Recall that in the leading order of the Newtonian approximation to 
general relativity, the $g_{00}$ component of the spacetime metric (in the 
natural gauge adapted to this approximation) is related to the Newton 
potential $\Phi$ via 
\be
g_{00}= -\left(1+\frac{1}{c^2}2\Phi+\ldots\right).
\ee
Comparing this to (\ref{metrel}), we find that our gauge field $A$ effectively 
plays the role of the Newton potential,
\be
A=-\Phi+\ldots.
\ee  
As we will see in Section~(\ref{compob}), this relationship is corrected by 
higher order terms already at the next order in the post-Newtonian 
approximation.  

\subsubsection{Extending the $1/c$ expansion}

We can also keep the subleading terms in the spatial metric, replacing
\be
g_{ij}\to g_{ij}-\frac{1}{c^2}\frac{A_{ij}(\bx,t)}{N}+\ldots
\ee
in (\ref{metrel}).  Following the rules of transformation for the spatial 
metric to one higher order in $1/c^2$ than before, it turns out that $A_{ij}$ 
also transforms under $\alpha$, 
\be
\delta_\alpha A_{ij}=\alpha\,\dot g_{ij}+N_i\nabla_j\alpha+N_j
\nabla_i\alpha.
\ee
This transformation property of $A_{ij}$ is just what is needed to remedy 
the noninvariance of our action under the local $U(1)$ transformations, by 
introducing a new coupling 
\be
\label{actaij}
\frac{2}{\kappa^2}\int dt\,d^D\bx\,\sqrt{g}\,A_{ij}\left(R^{ij}-\frac{1}{2}
Rg^{ij}\right).
\ee

Interestingly, this term is also ``accidentally'' invariant under another 
Abelian gauge symmetry, which acts on the fields via
\be
\label{varaij}
\delta A_{ij}=\nabla_i\varepsilon_j+\nabla_j\varepsilon_i.
\ee
The variation of all the other fields under the $\varepsilon_i$ symmetry 
is zero.  The total action is invariant under (\ref{varaij}):  The only term 
in the action which depends on $A_{ij}$ is (\ref{actaij}), and its invariance 
under (\ref{varaij}) is a consequence of the Bianchi identity.  

This new gauge symmetry (\ref{varaij}) also has a natural geometrical origin:  
In the process of decomposing the relativistic symmetries in the powers of 
$1/c$, we could have also kept the subleading terms in spatial diffeomorphisms, 
\be
\xi^i=\zeta^i-\frac{1}{c^2}\frac{\varepsilon^i(t,\bx)}{N}+\ldots.
\ee
The $c\to\infty$ limit of the relativistic diffeomorphisms then implies 
precisely the transformation rules (\ref{varaij}).

It thus appears that by extending the gravity multiplet to include both the 
Newton-potential $A(t,\bx)$ and the field $A_{ij}(t,\bx)$, we succeeded in 
finding a formulation of gravity in which the $U(1)\ltimes\diff(M,\CF)$ 
symmetry of ``nonrelativistic general covariance'' is realized in the full 
nonlinear theory without obstructions.  In addition, we have also seen that 
this extended gravity multiplet has a clear and natural geometric 
interpretation in the context of the $1/c$ expansion.   These features make 
this extended theory potentially attractive, but a closer inspection shows 
that the number of propagating degrees of freedom has been once again reduced 
to zero -- the theory turns out to be effectively topological.  
Consequently, the spectrum of bulk gravitons in the low-energy 
limit will not match the prediction of low-energy general relativity.  

In order to see this, and to count reliably the number of degrees of freedom, 
we turn once more to the Hamiltonian analysis \cite{ht}.   Because the 
subleading fields $A$ and $A_{ij}$ that we kept in the $1/c$ 
expansion appear in the action without time derivatives, they will all lead 
to constraints in the Hamiltonian formulation of the theory.  The full phase 
space is parametrized by fields $N_i$, $A$, $A_{ij}$, $g_{ij}$ and their 
canonical momenta $P^i$, $P_A$, $P^{ij}$ and $\pi^{ij}$, implying that 
\be
\label{aijdimp}
\dim\CP =2(D+1)^2
\ee
per spatial point.%
\footnote{There is also the canonical pair consisting of $N(t)$ and 
its conjugate momentum $P_0(t)$, which only yields an integral constraint 
and can be dropped for the purpose of counting the local degrees of freedom.}
The vanishing of the momenta conjugate to $A$, $A_{ij}$ and $N_i$ represents 
$(D +2)(D+1)/2$ primary constraints.  The condition that the primary 
constraints be preserved in time yields secondary constraints:  Insisting on 
$\dot P_A=0$ requires the vanishing of $R$, and similarly $\dot P^{ij}=0$ 
requires the vanishing of $R^{ij}-\frac{1}{2}Rg^{ij}$.  In addition, as in 
general relativity, $\dot P^i=0$ requires $\CH_i=0$.  

Naively, there are thus $D(D+3)/2+1$ secondary constraints $\CH_i$, $R$ and 
$R^{ij}-\frac{1}{2}Rg^{ij}$.  However, these are not all independent: 
$R^{ij}-\frac{1}{2}Rg^{ij}$ satisfies the Bianchi identity, and $R$ is 
proportional to the trace of $R^{ij}-\frac{1}{2}Rg^{ij}$, leaving $D(D+1)/2$ 
independent secondary constraints. 

All the primary and secondary constraints are first-class: Their commutators 
vanish on the constraint surface.  As a result, we have the total number
\be
\CC_1=(D+2)(D+1)/2+D(D+1)/2=(D+1)^2
\ee
of first-class constraints.  This implies, together with (\ref{aijdimp}) and 
invoking (\ref{ndof}), that the total number of local propagating degrees of 
freedom is
\be
\CN=\frac{1}{2}(\dim\CP-2\CC_1)=0.
\ee
The theory is effectively topological.  

Since our primary inerest in this paper is to find a theory whose spectrum 
of gravitons matches general relativity at long distances, we will not pursue 
the extended theory in which the gravity multiplet contains the $A_{ij}$ 
fields, and set $A_{ij}=0$ from now on.    

\subsection{Generally covariant nonrelativistic gravity in $2+1$ dimensions}
\label{sectwoone}

In $2+1$ dimensions, the Einstein tensor $R_{ij}-\frac{1}{2}Rg_{ij}$ of the 
spatial metric vanishes identically, which means that (\ref{ptwo}) is zero, 
and there is no obstruction against gauging the $\usigma$ symmetry in the full 
nonlinear theory.  The Noether procedure terminates after one step and leads 
to the following action, 
\be
\label{sa}
S=\frac{2}{\kappa^2}\int dt\,d^2\bx\,\sqrt{g}\left\{N\left(K_{ij}K^{ij}-
\lambda K^2-\CV\right)-AR\vphantom{g^{ik}}\right\}.
\ee
This action exhibits the $U(1)\ltimes\diff(M,\CF)$ gauge symmetry of 
nonrelativistic general covariance in $2+1$ dimensions, for any choice of 
$\CV$. 

The extended gauge symmetry eliminates the scalar degree of freedom of the 
graviton.  To see that, it is convenient to select $A=0$ as the gauge choice.  
In this gauge, the equations of motion are the same as in the minimal model 
with $\lambda=1$, with the addition of the Gauss constraint 
\be
\label{gauss}
R=0.
\ee
It is this additional constraint which eliminates the scalar degree of freedom 
of the minimal theory.  Moreover, since in $2+1$ dimensions the scalar 
graviton was the only local degree of freedom, the generally covariant theory 
with the extended $U(1)\ltimes\diff(M,\CF)$ symmetry has no local propagating 
graviton polarizations.  In this sense, it is akin to several other, much 
studied models of gravity in $2+1$ dimensions, such as standard general 
relativity or chiral gravity \cite{wss,mss}.  

Because of the absence of physical fluctuations, the geometry of classical 
solutions in this theory can be expected to be quite rigid, just as in the 
case of its relativistic cousins in $2+1$ dimensions.  In particular, the 
Gauss constraint (\ref{gauss}) forces the two-dimensional spatial slices to 
be flat.  It is natural to look for deformations of this theory which would 
at least replace the Gauss constraint with the more general condition of 
constant spatial curvature, but there appear to be no consistent deformations 
that could modify the Gauss constraint to 
\be
R=2\Omega,
\ee
 with $\Omega$ a new coupling constant.  However, once we learn in 
Section~\ref{secg} how to gauge the $\usigma$ symmetry in the general case of 
$D+1$ dimensions, we will also find a mechanism for turning on this new 
coupling $\Omega$.  

\subsection{Self-interacting Abelian gravity}

Another way to eliminate the obstruction against gauging $\usigma$ is to 
linearize the gauge symmetries of the minimal theory.  The fields in the 
theory with linearized gauge symmetries are $h_{ij}$, $n_i$ and $n$.  The 
gauge transformations $\xi_i(t,\bx)$ and $f(t)$ act via
\bea
\label{trsfl}
\delta h_{ij}&=&\p_i\xi_j+\p_j\xi_i,\nonumber\\
\delta n_i&=&\dot\xi_i,\\
\delta n&=&\dot f,\nonumber
\eea
and represent the Abelian contraction of $\diff(M,\CF)$.  

The kinetic term (with $\lambda=1$) takes the form
\be
\label{kinlin}
S_K=\frac{1}{2}\int dt\,d^D\bx\left(\dot h_{ij}-\p_in_j-\p_jn_i\right)
\left(\delta_{ik}\delta_{j\ell}-\delta_{ij}\delta_{k\ell}\right)
\left(\dot h_{k\ell}-\p_kn_\ell-\p_\ell n_k\right).
\ee
In this theory, the obstruction (\ref{ptwo}) against gauging vanishes 
identically, as was already established in our analysis of the linearized 
approximation to the minimal theory in Section~\ref{secsy}.  

At first glance, it would thus seem that keeping only the linearized gauge 
symmetries would reduce the model to the nonintreracting Gaussian theory 
studied in Section~\ref{secsy}, but in fact it is not so.  Even though the 
kinetic term takes the Gaussian form (\ref{kinlin}), the potential term 
need not be Gaussian.  

Suitable terms in $\CV$ are integrals of local operators, which are either 
invariant under (\ref{trsfl}), or invariant up to a total spatial derivative.  
The building blocks that can be used to construct such operators are the 
linearized curvature tensor of the spatial metric, and its derivatives.  
We will denote the linearized Riemann tensor by
\be
L_{ijk\ell}=\frac{1}{2}\left(\p_j\p_k h_{i\ell}-\p_j\p_\ell h_{ik}-\p_i\p_k h_{j\ell} 
+\p_i\p_\ell h_{jk}\right), 
\ee
and similarly the linearized Ricci tensor by $L_{ij}\equiv L_{ikjk}$ and the 
Ricci scalar by $L\equiv L_{ii}$.  Clearly, there is an infinite hierarchy of 
sutable operators, which reduces to a finite number if we limit the number of 
spatial derivatives by $2z$.  For interesting values of $z>1$, the general 
$\CV$ built from such terms will not be purely Gaussian, leading to a 
self-interacting theory.  

Thus, in the context of gravity with anisotropic scaling, linearizing the 
gauge symmetries does not necessarily make the theory noninteracting -- 
we find a novel interacting theory of Abelian gravity instead.  Curiously, 
a similar phenomenon has been observed in the case of general relativity 
25 years ago by Wald \cite{wald}, where it was shown that by taking the 
action to contain higher powers of the linearized curvature, one can 
construct a self-interacting theory of spin-two fields in flat spacetime 
with linearized spacetime diffeomorphisms as gauge symmetries.  In the 
relativistic case studied in \cite{wald}, this construction leads inevitably 
to higher time derivatives in the action, and therefore problems with ghosts 
in perturbation theory.  In contrast, our nonrelativistic models of 
self-interacting Abelian gravity do not suffer from this problem -- their 
self-interaction results from higher than quadratic terms in $\CV$, with the 
kinetic term taking the Gaussian form (\ref{kinlin}).  For suitable choices 
of the couplings in $\CV$, the spectrum is free of both ghosts and tachyons.

For an arbitrary $\CV$, the gauging of $\usigma$ is now accomplished by adding 
a new Gaussian term to the action, 
\be
-2\int dt\,d^D\bx\,AL.
\ee
The theory is now gauge invariant under the linearized action of $U(1)$, 
\be
\delta A=\dot\alpha,\qquad\delta n_i=\p_i\alpha.
\ee
Arguments identical to those in Section~\ref{secsy} show that the theory 
contains only the tensor graviton modes, eliminating the scalar.  

At long distances, the dominant terms in $\CV$ are those with the lowest 
number of spatial derivatives.  Since the only suitable operator with 
just two derivatives is the quadratic part of the spatial Einstein-Hilbert 
term, 
\be
\label{lineh}
\int dt\,d^D\bx\,\left(h_{ij}L_{ij}-\frac{1}{2}h_{ii}L\right),
\ee
the theory becomes automatically Gaussian at long distances, and approaches a 
free infrared fixed point with $z=1$.  This behavior can be avoided if we 
insist that all operators $\CV$ are integrals of gauge-invariant operators:  
Since (\ref{lineh}) is only invariant up to a total derivative, it does not 
belong to this class.  In these restricted theories, the infrared behavior 
will be controlled by Gaussian terms with $z\geq 2$.  In fact, such 
self-interacting Abelian gravity theories, approaching free-field 
Lifshitz-type fixed points $z\geq 1$, have been encountered in the infrared 
regime of a family of condensed matter models on the rigid fcc lattice in 
\cite{eme}.

While such self-interacting Abelian gravity models might be useful for 
describing new universality classes of bose liquids in condensed matter 
theory, they do not appear phenomenologically viable as candidates for 
describing the gravitational phenomena in the observed universe.    

\section{General Covariance at a Lifshitz Point}
\label{secgcl}

So far, we focused on the special cases in which the obstruction to the 
gauging of $\usigma$ is absent.  However, none of the resulting theories of 
gravity with extended gauge symmetry discussed in Section~\ref{secga} appear 
phenomenologically interesting as models of gravity in $3+1$ dimensions.  

Here we change our perspective, and present a robust mechanism which 
allows $\usigma$ to be gauged in the general spacetime dimension $D+1$.  
This will lead to a theory of gravity with nonrelativistic general covariance 
which reproduces many properties of general relativity at long distances.  

\subsection{Repairing the global $\usigma$ symmetry}
\label{secrep}

In Section~\ref{secnonl}, we found an obstruction that prevents the 
$\usigma$ from being a global symmetry of the full nonlinear theory 
for $D>2$.  Leaving aside the possibility that this obstruction could be 
cancelled by quantum effects (perhaps by a mechanism similar to 
\cite{hw1,hw2}), we look for a way to repair the $\usigma$ symmetry at the 
classical level.  

\subsubsection{The Newton prepotential}

In order to eliminate the obstruction, we introduce an auxiliary scalar field 
$\nu$, which transforms under $\usigma$ as
\be
\delta_\alpha\nu=\alpha.
\ee
We will refer to this field as the ``Newton prepotential.''  The scaling 
dimension of $\nu$ is the same as the dimension of $\alpha$, 
\be
[\nu]=z-2.
\ee

We can now repair the $\usigma$ symmetry by adding a new term to the action, 
\bea
S_\nu&=&\frac{2}{\kappa^2}\int dt\,d^D\bx\,\sqrt{g}\,\nu\left(R^{ij}
-\frac{1}{2}Rg^{ij}\right)\left(\dot g_{ij}-\nabla_iN_j-\nabla_jN_i\right)
\nonumber\\
&&\qquad{}+\frac{2}{\kappa^2}\int dt\,d^D\bx\,\sqrt{g}\,N\,\nu\left(R^{ij}
-\frac{1}{2}Rg^{ij}\right)\nabla_i\nabla_j\nu.
\label{newtm}
\eea
The variation of $\nu$ in the linear term compensates for the noninvariance 
of the original action of the minimal theory.  The term quadratic in $\nu$ 
is in turn required to cancel the variation of $N_i$ in the term linear in 
$\nu$.  

\subsubsection{Relevant deformations}

We can check by linearizing around the flat background that the number of 
propagating degrees of freedom has not changed by the introduction of the 
Newton prepotential terms in the action.  It is rather unconventional that 
in the expansion around the flat spacetime, the new field $\nu$ enters the 
action at the cubic order in small fields, \ie , its presence does not affect 
the propagator.  This issue is eliminated by noticing that a new term, of 
lower dimension and also invariant under the global $\usigma$ symmetry, can 
be added to the action:
\be
\label{somega}
S_\Omega=\frac{2\Omega}{\kappa^2}\int dt\,d^D\bx\,\sqrt{g}\,\nu g^{ij}\left(
\dot g_{ij}-\nabla_iN_j-\nabla_jN_i\right)+\frac{2\Omega}{\kappa^2}\int dt\,
d^D\bx\,\sqrt{g}\,N\,\nu\Delta\nu.
\ee
Here $\Omega$ is a coupling constant of dimension $[\Omega]=2$.    
With the addition of this relevant term, the Newton prepotential enters 
the linearized theory, at the quadratic order in fields around the flat 
spacetime. 

The Newton prepotential enters the action with global $\usigma$ symmetry 
quadratically, and can be integrated out by solving its equation of motion,
\be
\label{nem}
\Theta^{ij}\nabla_i\nabla_j\nu+\Theta^{ij}K_{ij}=0,  
\ee
where we have introduced 
\be
\Theta^{ij}=R^{ij}-\frac{1}{2}Rg^{ij}+\Omega g^{ij}.
\ee
Integrating out $\nu$ would result in a nonlocal action, because (\ref{nem}) 
is solved by
\be
\label{nuzero}
\nu_0(\bx,t)=-\int d^D\bx'\frac{1}{\Theta^{ij}\nabla_i\nabla_j}(\bx,
\bx')(\Theta^{k\ell}K_{k\ell})(\bx').
\ee
Here we have assumed that the operator $\Theta^{ij}\nabla_i\nabla_j$ is 
invertible, and denoted its Green's function by 
$(\Theta^{ij}\nabla_i\nabla_j)^{-1}(\bx,\bx')$.  We will not try to determine 
the exact conditions under which this assumption is true.  However, for 
example near the flat spacetime geometry (\ref{flatst}), we have 
$\Theta^{ij}\nabla_i\nabla_j=\Omega\Delta+\CO(h_{ij})$, where $\Delta=\p^2$ 
is the flat-space Laplacian.  This operator is invertible at least in 
perturbation theory, as long as we keep $\Omega$ nonzero.%
\footnote{Another interesting example, which will be important below when we 
gauge the $\usigma$ symmetry, is the case in which $\hat g_{ij}$ is the metric 
of a maximally symmetric space satisfying $\hat R=2\Omega$.  In this reference 
background, we have $\Theta^{ij}\nabla_i\nabla_j=2\Omega\hat\Delta/D$ (with 
$\hat\Delta$ the Laplace operator of $\hat g_{ij}$), which is also invertible 
for $\Omega\neq 0$.}

Note that the Green's function is still a local function in $t$.  
Note also that the expression (\ref{nuzero}) for $\nu_0$ has the right form in 
order for the action to be $\usigma$ invariant after $\nu$ has been integrated 
out.  In particular, 
\bea
\delta_\alpha\nu_0(\bx,t)&=&-\frac{1}{2}\int d^D\bx'\frac{1}{\Theta^{ij}
\nabla_i\nabla_j}(\bx,\bx')\Theta^{k\ell}(-\nabla_k\nabla_\ell\alpha
-\nabla_\ell\nabla_k\alpha)(\bx')\cr
&&\qquad{}=\int d^D\bx'\frac{1}{\Theta^{ij}\nabla_i\nabla_j}
(\bx,\bx')(\Theta^{k\ell}\nabla_k\nabla_\ell\alpha)(\bx')=\alpha(\bx,t).
\eea
The nonlocality of the action obtained by integrating out $\nu$ is 
relatively mild:  In particular, this nonlocality is purely spatial, along the 
leaves of the spacetime foliation $\CF$.  Such nonlocalities are quite common 
in rather conventional condensed matter systems.  Nevertheless, in the rest of 
the paper, we will keep the action manifestly local, by keeping the Newton 
prepotential $\nu$ as an independent field instead of integrating it out.  

\subsubsection{Linearized theory with global $\usigma$ around flat spacetime}
\label{seclin}

First we will check that our repair of the global $\usigma$ symmetry has 
not changed the count of the number of degrees of freedom.  
Even with the $\Omega$ coupling turned on, the flat spacetime geometry 
\be
g_{ij}=\delta_{ij},\qquad N_i=0,\qquad N=1,\qquad \nu=0
\ee
is still a classical solution of the theory (with $\Lambda=0$), and we can 
expand around it.  

The $\nu$ equation of motion is
\be
\label{linnueom}
2\p^2(\nu-C)+\dot h=0.
\ee
The momentum constraints give
\be
\label{linmomi}
\p^2(\dot w_i-u_i)=0
\ee
and
\be
\label{linmomc}
2\Omega\p^2\nu+\frac{D-1}{D}\left\{(\p^2)^2\dot B-\p^2\dot h\right\}=0.
\ee
Setting $B=0$ and $w_i=0$ fixes gauge completely, and implies (with 
appropriate boundary conditions at infinity) that $u_i=0$ and 
\be
\label{solnnu}
2\Omega\nu=\frac{D-1}{D}\dot h.
\ee
In this gauge, the remaining equations of motion are
\be
-\ddot s_{ij}+\frac{D-1}{D}\delta_{ij}\ddot h+2(\p_i\p_j-\delta_{ij}\p^2)\dot C
-2\Omega\delta_{ij}\dot\nu-\frac{\delta\CV_2}{\delta g_{ij}}=0,
\ee
where $\CV_2$ denotes the quadratic part of $\CV$ in the linearized theory.  
Using (\ref{solnnu}), we see that the $\dot\nu$ term cancels the $\ddot h$ 
term exactly, allowing one to determine $h$ as a function of $\dot C$, and 
substitute back into (\ref{linnueom}).  The resulting equation determines 
the dispersion relation of the scalar polarization of the graviton.  For 
example, when we set the cosmological constant $\Lambda=0$, the potential term 
will be dominated at long distances by $\CV=-R$, and we get 
\be
\frac{\delta\CV_2}{\delta g_{ij}}=-\p^2s_{ij}+\frac{D-2}{D}(\p_i\p_j
-\delta_{ij}\p^2)(\p^2B-h).
\ee
The metric equation of motion then implies that 
\be
h=\frac{2D}{2-D}\dot C,
\ee
and the $\nu$ equation of motion gives
\be
\label{ceom}
\frac{D-1}{\Omega}\p^2\ddot C+D\ddot C+(D-2)\p^2C=0.
\ee
The spectrum thus contains the transverse, traceless polarizations $s_{ij}$ 
with dispersion 
\be
\label{disptt}
\omega^2=\bk^2,
\ee
plus a scalar graviton (described in this gauge by $C$) which exhibits the 
dispersion relation implied by (\ref{ceom}), 
\be
\label{dispsc}
\omega^2=-\frac{(D-2)\bk^2}{\displaystyle{D\left\{
1-\frac{D-1}{D\Omega}\bk^2\right\}}}.
\ee

Note that the scalar mode is inevitably tachyonic at low energies.  
This is implied by the choice of sign in $\CV$, determined from the 
requirement that the tensor polarizations have the correct-sign dispersion 
relation (\ref{disptt}).  This tachyonic nature of the scalar mode is not 
a cause for any concern, because the model discussed here represents only an 
intermediate stage of our construction -- our intention is to gauge the 
$\usigma$ symmetry, which will turn the scalar graviton into a gauge artifact.

Note also that taking the regulating dimensionful coupling $\Omega$ to 
zero reduces the scalar dispersion relation (\ref{dispsc}) correctly to the 
singular dispersion $\omega^2=0$, observed at $\lambda=1$ in the minimal 
theory in \cite{mqc} and in (\ref{dispone}).

\subsection{Gauging the $\usigma$ symmetry}
\label{secg}

Having repaired the global $\usigma$ symmetry, we can now gauge it.  The 
Noether method closes after just one step; adding
\be
\label{noet}
S_{A,\Omega}=-\frac{2}{\kappa^2}\int dt\,d^D\bx\,\sqrt{g}\,A\,(R-2\Omega)
\ee
to the action makes the theory gauge invariant under the $U(1)$ symmetry.  
This procedure results in the following action of the generally covariant 
theory of gravity with anisotropic scaling,   
\be
\label{fullact}
S=\frac{2}{\kappa^2}\int dt\,d^D\bx\,\sqrt{g}\left\{N\left[K_{ij}K^{ij}-K^2-\CV
+\nu\,\Theta^{ij}\vphantom{g^{ik}}\left(2K_{ij}+\nabla_i\nabla_j\nu\right)
\right]-A\,(R-2\Omega)\right\},  
\ee
with $\Theta^{ij}$ a short-hand notation for
\be
\Theta^{ij}=R^{ij}-\frac{1}{2}g^{ij}R+\Omega g^{ij}.
\ee

Note that in the theory with the Newton prepotential $\nu$, the issue about 
the possibility of adding the spatial cosmological constant $\Omega$, raised in 
the generally covariant theory in $2+1$ dimensions at the end of 
Section~\ref{sectwoone}, has been resolved by the introduction of the Newton 
prepotential.  

Note also that in addition to the newly introduced gauge field $A$, the theory 
contains a composite 
\be
a=\dot\nu-N^i\nabla_i\nu+\frac{N}{2}\nabla^i\nu\nabla_i\nu
\ee
which also transforms as a gauge field under the $U(1)$ gauge transformations,
\be
\delta_\alpha a=\dot\alpha-N^i\nabla_i\alpha.
\ee
Moreover, both $A$ and the composite gauge field $a$ share the same 
transformation properties under the rest of the gauge group,
\be
\delta a=\xi^i\p_i a+\dot f a+f\dot a.
\ee
As a result, the gauged action stays gauge invariant if we replace 
\be
\label{subst}
A\to (1-\gamma)A+\gamma a,
\ee
with $\gamma$ a real coefficient.  

In fact, the composite field $a$ already made its appearance in the theory 
with the global $\usigma$ symmetry presented in Section~\ref{secrep}:  Up to a 
total derivative, the relevant term (\ref{somega}) can be rewritten as
\be
S_\Omega=-\frac{4\Omega}{\kappa^2}\int\sqrt{g}\,a.
\ee
Hence, the substitution (\ref{subst}) in the $\int dt\,d^D\bx\,\sqrt{g}A$ term 
in (\ref{noet}) will just shift the effective value of $\Omega$.  Similarly, 
substituting (\ref{subst}) in the $AR$ term in the Noether coupling 
(\ref{noet}) effectively shifts the coefficients in (\ref{newtm}).  

\subsubsection{Hamiltonian formulation}

The structure of gauge symmetries can be verified by analyzing 
the algebra of Hamiltonian constraints of the theory.  In addition, this 
analysis will allow us to obtain the precise count of the number of 
propagating degrees of freedom, using formula (\ref{ndof}).  This approach 
to the count of the degrees of freedom is usually more accurate and more 
reliable than our previous analysis of the linearized spectrum around a fixed 
solution, for at least two reasons.  First, it is less background-dependent, 
because it sidesteps the need to linearize the theory around a fixed 
solution.  Secondly, because it is valid for the full nonlinear theory, 
it excludes the possible artifacts of the linearized approximation.  

We will set $\kappa=1$ to eliminate additional clutter, and denote the 
canonical momenta conjugate to the spatial metric by $\Pi^{ij}$: 
\be
\Pi^{ij}=\frac{\delta S}{\delta \dot g_{ij}}=2\sqrt{g}\left(K^{ij}-g^{ij}K
+\Theta^{ij}\nu\right)=\pi^{ij}+2\sqrt{g}\,\Theta^{ij}\nu.  
\ee
The lower-case $\pi^{ij}$ are reserved for the standard expressions 
for the canonical momenta in general relativity, 
\be
\pi^{ij}\equiv 2\sqrt{g}\left(K^{ij}-g^{ij}K\right).
\ee
The remaining canonical momenta
\be
P^i(\bx,t)\equiv\frac{\delta S}{\delta\dot N_i},\qquad 
p_\nu(\bx,t)\equiv\frac{\delta S}{\delta\dot\nu},\qquad
P_A(\bx,t)\equiv\frac{\delta S}{\delta\dot A},\qquad
P_0(t)\equiv\frac{\delta S}{\delta\dot N}
\ee
all vanish, and represent the primary constraints.  The Poisson brackets are
\bea
[g_{ij}(\bx,t),\Pi^{k\ell}(\by,t)]&=&\frac{1}{2}\left(\delta^k_i\delta^\ell_j+
\delta^\ell_i\delta^k_j\right)\delta(\bx-\by),\cr
[N_i(\bx,t),P^j(\by,t)]&=&\delta^j_i\delta(\bx-\by),\qquad\qquad\qquad\!
[N(t),P_0(t)]=1,\cr
[A(\bx,t),P_A(\by,t)]&=&\delta(\bx-\by),\qquad\qquad\quad
[\nu(\bx,t),p_\nu(\by,t)]=\delta(\bx-\by),\nonumber
\eea
and zero otherwise.  

In the canonical variables, the Hamiltonian is given by
\bea
H&=&\int d^D\bx\,\left\{N\left[\frac{1}{2\sqrt{g}}\left(\Pi^{ij}-2\sqrt{g}\,
\Theta^{ij}\nu\right)\CG_{ijk\ell}\left(\Pi^{k\ell}-2\sqrt{g}\,\Theta^{k\ell}
\nu\right)\right.\right.\nonumber\\
&&\qquad\qquad\left.\left.{}+2\sqrt{g}\,\Theta^{ij}\nabla_i\nu
\nabla_j\nu+2\sqrt{g}\,\CV\vphantom{\frac{1}{2}}\right]
-2N_i\nabla_j\Pi^{ij}+2\sqrt{g}\,A\,(R-2\Omega)\vphantom{\frac{1}{2}}\right\},
\eea
where 
\be
\CG_{ijk\ell}=\frac{1}{2}\left(g_{ik}g_{j\ell}+g_{i\ell}g_{jk}\right)
-\frac{1}{D-1}g_{ij}g_{k\ell}
\ee
is the inverse of the De~Witt metric $G^{ijk\ell}$ of (\ref{dwm}) for 
$\lambda=1$. 

At this stage, the primary constraints are included in the Hamiltonian with 
the use of Lagrange multipliers $U_i(\bx,t)$, $\CU(\bx,t)$, $U_A(\bx,t)$, and 
$U_0(t)$, 
\be
\label{hamu}
H\to \hat H=H+\int d^D\bx\left(U_iP^i+\CU p_\nu+U_AP_A\right)+U_0P_0.
\ee
The preservation of the primary constraints under the time evolution given 
by (\ref{hamu}) requires that the commutators of the primary constraints with 
$\hat H$ vanish, yielding the following set of secondary constraints which are 
local in space,
\bea
\CH^i&\equiv&[\hat H,P^i]=-2\nabla_j\Pi^{ij},\\
\Phi&\equiv&[\hat H,p_\nu]=-4\sqrt{g}\,N\,\Theta^{ij}\nabla_i\nabla_j\nu
+4\sqrt{g}\,N\,\Theta^{ij}\CG_{ijk\ell}\Theta^{k\ell}\nu-2N\,\Theta^{ij}\CG_{ijk\ell}
\Pi^{k\ell},\\
\Psi&\equiv&[\hat H,P_A]=2\sqrt{g}{(R-2\Omega)},
\eea
and one integral constraint
\bea
\label{intc}
\int d^D\bx\,\CH_0&\equiv&[\hat H,P_0]=\int d^D\bx\,\left\{\frac{1}{2\sqrt{g}}
\left(\Pi^{ij}-2\sqrt{g}\,\Theta^{ij}\nu\right)\CG_{ijk\ell}\left(\Pi^{k\ell}
-2\sqrt{g}\,\Theta^{k\ell}\nu\right)\right.\nonumber\\
&&\qquad\qquad\qquad\qquad\left.{}+2\sqrt{g}\,\Theta^{ij}\nabla_i\nu\nabla_j\nu
+2\sqrt{g}\,\CV\vphantom{\frac{1}{2}}\right\}.
\eea
This integral constraint will not affect the number of local degrees of 
freedom.  To avoid unnecessary clutter, we concentrate on the analysis of the 
local constraints, returning to (\ref{intc}) only at the end of this section.

Next, we need to ensure that the secondary constraints are preserved in time.  
The momentum constraints $\CH^i$ take formally the same form as in general 
relativity or in the minimal theory of \cite{mqc,lif}.  They are indeed 
preserved in time, albeit in a slightly more intricate way than in the minimal 
theory or in general relativity.  In those cases (see the discussion in 
Section~4.4 of~\cite{mqc}), the commutator of $\CH^i$ with $H$ only gets 
a contribution from the $N_k\CH^k$ terms in $H$.  The rest of the commutator 
between the density of the Hamiltonian and $\CH^i$ adds up to a total 
derivative, as a consequence of the transformation properties of a scalar 
density under spatial diffeomorphisms.  Here, the argument is more subtle, 
and the commutator contains additional terms, 
\be
[\hat H,\CH^i]=-\nabla_k(N^k\CH_i)-(\nabla_iN^k)\CH_k-(\nabla^i\nu)\Phi
-(\nabla^i A)\Psi.  
\ee
However, this expression vanishes on the constraint surface, and no tertiary 
constraints are produced at this stage.  

The time preservation of the secondary constraint $\Phi$ requires the 
vanishing of
\be
\label{tprsec}
[\hat H,\Phi]\equiv 4\sqrt{g}\,N\,\Theta^{ij}\left(\nabla_i\nabla_j-\CG_{ijk\ell}
\Theta^{k\ell}\right)\CU+[H,\Phi]-U_0\frac{\Phi}{N}=0.
\ee
Unlike the conditions for the time preservation of the primary constraints 
or the $\CH^i$, condition (\ref{tprsec}) depends explicitly on one of the 
Lagrange multipliers, $\CU$.  Therefore, setting $[\hat H,\Phi]=0$ yields an 
equation for $\CU$, instead of producing an additional, tertiary constraint.  
Also, because the commutator 
\be
\label{pphicom}
[p_\nu(\bx),\Phi(\by)]=4\sqrt{g}\,N\,\Theta^{ij}\left(\nabla_i\nabla_j
-\CG_{ijk\ell}\Theta^{k\ell}\right)\delta(\bx-\by)
\ee
does not vanish on the constraint surface, $p_\nu$ and $\Phi$ represent a pair 
of second-class constraints.  

It now remains to check the condition for the preservation of $\Psi$ in time.  
After a lengthy calculation, we get
\be
[\hat H,\Psi]=+N\nabla_i\CH^i-\Phi-\nabla_i\left(N^i\Psi\right).
\ee
This expression vanishes on the constraint surface.  Again, no tertiary 
constraint is produced, and the process of generating the full list of 
constraints stops here.  

One might be tempted to expect that $\Psi$ is a first-class constraint, but 
that expectation is false:  The commutator of $\Psi(\bx)$ and $\Phi(\by)$ 
does not vanish on the constraint surface.  Consequently, the first-class 
and second-class constraints are still entangled, and $\Psi(\bx)$ itself 
is a mixture of constraints of both classes.  In order to disentangle the 
constraints, we must first evaluate
\bea
[\Psi(\bx),\Phi(\by)]&=&-4
\frac{\displaystyle{\delta\left\{\sqrt{g}(R-2\Omega)(\bx)\right\}}}{
\displaystyle{\delta g_{ij}(\by)}}\left(N\,\CG_{ijk\ell}\Theta^{k\ell}\right)
(\by)\nonumber\\
&=&4\sqrt{g}\,N\,\Theta^{ij}\left(\CG_{ijk\ell}\Theta^{k\ell}
-\nabla_i\nabla_j\right)\delta(\bx-\by).
\eea
This is equal, up to a sign, to the commutator of $p_\nu$ and $\Phi$ which we 
obtained in (\ref{pphicom}).  Hence, it is natural to define
\be
\CH_A=\Psi+p_\nu.  
\ee
$\CH_A$ commutes both with $\Phi$ and with $p_\nu$, and represents a 
first-class constraint.  

Having identified $\CH_A$ as the final first-class constraint, we can check 
that it generates the correct $U(1)$ gauge transformations on the fields.  
In the Hamiltonian formalism, the gauge transformation generated by 
a first-class constraint on an arbitrary phase-space variable $\phi$ is given 
by the commutator of $\phi$ with the corresponding constraint \cite{ht}, 
for example
\be
\delta_\alpha\phi(\bx,t)=-[\int d^D\by\,\alpha(\by,t)\CH_A,\phi(\bx,t)].
\ee
One can indeed use this Hamiltonian formula to check that the gauge symmetries 
implied by the first-class constraints reproduce those that we found in the 
Lagrangian formulation above.  

Given our analysis of the constraints, we can now evaluate the number of 
degrees of freedom.  Altogether, the theory has $\dim\CP=D^2+3D+4$ 
canonical variables per spacetime point.  These variables are constrained by 
$\CC_1=2D+2$ first-class constraints ($P^i$, $P_A$, $\CH^i$ and 
$\CH_A$), and $\CC_2=2$ second-class constraints ($p_\nu$ and $\Phi$).  
The number of degrees of freedom $\CN$ per spacetime point is then given by 
formula (\ref{ndof}),
\be
\label{ngc}
\CN=\frac{1}{2}\left(\dim\CP-2\CC_1-\CC_2\right)=
\frac{1}{2}(D+1)(D-2).
\ee
This correctly reproduces the number of tensor (\ie\ transverse, traceless) 
polarizations of the graviton in $D+1$ spacetime dimensions.  

Returning to the integral constraint (\ref{intc}), we note that its 
commutation relations with the rest of the constraint algebra can be read 
off from the commutators of $H$ obtained above.  This follows from the fact 
that, as in general relativity, the Hamiltonian can be written as a sum of 
constraints,
\be
H=N\int d^D\bx\,\CH_\perp+\int d^D\bx\left(N^i\CH_i+A\Psi\right).
\ee

Actually, the role of the integral constraint (\ref{intc}) deserves to be 
investigated further.  It is plausible that in the theory with nonrelativistic 
general covariance, where the $U(1)$ gauge symmetry mimics the role of 
relativistic time reparametrizations, one can choose not to impose the 
integral constraint on physical states.  This would be eqiuvalent to the 
omission of nonrelativistic time reparametrizations $\delta t=f(t)$ 
from the gauge symmetries, effectively setting $N(t)=1$.   If consistent, 
this construction would lead to a theory of gravity with nonzero energy levels 
even in spacetimes with compact spatial slices $\Sigma$.  In fact, this 
situation was already encountered on flat noncompact $\Sigma$ in the context 
of Abelian gravity in \cite{eme}.  On noncompact $\Sigma$, the possibility of 
relaxing the integral Hamiltonian constraint will be closely tied to the 
structure of consistent boundary conditions at infinity in gravity with 
anisotropic scaling, whose study has been initiated in \cite{aci}.

\subsubsection{Linearization around detailed balance}

In principle, our result (\ref{ngc}) for the number of degrees of freedom 
$\CN$ can be checked by linearizing the theory around a chosen solution, and 
explicitly counting the number of propagating polarizations.  However, in 
order to investigate the spectrum of the linearized theory after gauging, we 
cannot use the flat spacetime as a reference background, because it no longer 
solves the equations of motion if $\Omega$ is not zero.  

This makes the analysis of the linearized approximation for general values 
of the couplings algebraically tedious, and we will not present it here in 
full generality.  Instead, we content ourselves with testing (\ref{ngc}) in 
the simpler case when the theory satisfies the detailed balance condition.  
Hence, we assume that the potential takes the special form 
\be
\CV=\frac{1}{4}\CG_{ijk\ell}\frac{\delta W}{\delta g_{ij}}
\frac{\delta W}{\delta g_{k\ell}},
\ee
and for concreteness we choose 
\be
W=\frac{1}{2\kappa_W^2}\int d^D\bx\,\sqrt{g}(R-2\Lambda_W^{}).
\ee
Before the $\usigma$ is gauged, the theory in detailed balance admits a 
particularly simple static ground-state solution,
\be
g_{ij}=\hat g_{ij}(\bx),\qquad N=1,\qquad N_i=0,\qquad\nu=0,
\ee
where $\hat g_{ij}$ is the maximally symmetric spatial metric which solves 
the equations of motion of $W$, 
\be
R_{ij}-\frac{1}{2}R g_{ij}+\Lambda^{}_Wg_{ij}=0.
\ee
In order for this background to be a solution of the theory with the 
extended $U(1)\ltimes\diff(M,\CF)$ gauge symmetry, we must set the spatial 
cosmological constant $\Omega$ equal to
\be
\Omega=\frac{D}{D-2}\Lambda^{}_W.
\ee
For $\Omega>0$, the ground-state geometry is the Einstein static universe, 
with spatial slices $\Sigma=S^D$.  Conversely, when $\Omega<0$, the ground 
state is the hyperbolic version of the Einstein static universe, with 
noncompact $\Sigma$.  Its curvature tensor satisfies 
$\hat R_{ij}=\frac{2\Omega}{D}\hat g_{ij}$ and $\hat R=2\Omega$.  

We now determine the spectrum of linearized perturbations around this class 
of ground state solutions.  The analysis closely parallels that of 
Sections~\ref{secsy} and \ref{seclin}, and we will be brief.  We expand the 
metric,  $g_{ij}=\hat g_{ij}+\kappa h_{ij}$, and decompose the linearized 
fluctuations as in (\ref{irrflg}) and (\ref{irrfln}):
\bea
h_{ij}&=&s_{ij}+\hat\nabla_iw_j+\hat\nabla_jw_i+\left(\hat\nabla_i\hat\nabla_j-
\frac{1}{D}\hat g_{ij}\hat\Delta\right)B+\frac{1}{D}h\,\hat g_{ij},\cr
n_i&=&u_i+\hat\nabla_iC,
\eea
with $\hat\nabla_i$ the covariant derivative of $\hat g_{ij}$.  The $\nu$ 
equation of motion is
\be
\label{nueom}
\frac{2\Omega}{D}\left(\hat\Delta\nu+\frac{1}{2}\dot h-\hat\Delta C\right)=0,
\ee
and the momentum constraints give
\bea
\left(\hat\Delta+\frac{2\Omega}{D}\right)(\dot w_i-u_i)&=&0,\cr
\hat\nabla_i\left(\frac{2\Omega}{D}\dot B+\frac{D-1}{D}(\hat \Delta B-\dot h)
+\frac{4\Omega}{D}\nu\right)&=&0.
\eea

To fix the $\diff(M,\CF)$ symmetries, we set $w_i=B=n=0$.  In this gauge, the 
momentum constraints 
reduce to
\be
(\hat\Delta +2\Omega/D)u_i=0,\qquad \hat\nabla_i\left[4\Omega\,\nu-(D-1)
\dot h\right]=0,
\ee  
which implies, with suitable boundary conditions, that $u_i$ is not 
propagating, and that 
\be
\Omega\,\nu=\frac{D-1}{4}\dot h.
\ee
Plugging this back into (\ref{nueom}) yields
\be
\label{nueomr}
\frac{D-1}{4\Omega}\hat\Delta\dot h+\frac{1}{2}\dot h-\hat\Delta C=0.
\ee

Finally, there is the constraint $R-2\Omega=0$, which plays the role of the 
Gauss constraint in our gauge $A=0$.  Its linearization around our detailed 
balance background 
\be
\label{gslin}
R-\hat R\approx-\frac{1}{D}\left[(D-1)\hat\Delta h+2\Omega h\right]=0
\ee
shows that $h$ is not propagating.  Combining (\ref{gslin}) with 
(\ref{nueomr}) then implies that $\hat\Delta C=0$.  Hence, the only propagating 
modes are the transverse traceless polarizations of the graviton $s_{ij}$.  
In particular, the scalar graviton has been eliminated, and the number of 
physical degrees of freedom agrees with the result of our Hamiltonian 
analysis (\ref{ngc}).  

\section{Conclusions}

In this paper, we have found a formulation of the theory of gravity with 
anisotropic scaling in which the gauge symmetry of foliation-preserving 
diffeomorphisms $\diff(M,\CF)$ is enhanced to the symmetry of 
``nonrelativistic general covariance,'' $U(1)\ltimes\diff(M,\CF)$.  

The advantage of this construction 
is that it relies only on the structure of the kinetic term in the action 
(\ref{smint}) (and, in fact, forces it to take the general-relativistic form 
with $\lambda=1$), while the form of the potential term $\CV$ is left 
unconstrained.  Therefore, we can consider the scenario proposed originally 
in \cite{mqc,lif}, in which the theory is defined at short distances by 
a $z>1$ fixed point (with $\CV$ dominated by higher-derivative terms), and 
is then expected to flow under the influence of relevant terms to $z=1$ and 
isotropic scaling in the infrared.  This classical scenario will of course 
receive quantum corrections, which could drive the theory outside the range 
of validity of the covariant action (\ref{fullact}).  In the rest of the 
paper, we limit our attention to the possibility that the long-distance 
physics is still described by the same action (\ref{fullact}), with $\CV$ 
dominated by the most relevant terms (\ref{vir}).

The first good news is that, as a result of the extended gauge symmetry, the 
spectrum contains just the transverse-traceless (tensor) modes of the 
graviton.  The scalar graviton mode of the minimal theory has been 
eliminated.  In $3+1$ spacetime dimensions, the elimination of the scalar mode 
has an interesting consequence in the short-distance regime of the theory.  
Recall that in the minimal theory with the potential dominated at short 
distances by the $z=3$ term (\ref{vuv}), the scalar mode is the sole physical 
mode that does not get a contribution to its 
dispersion relation from (\ref{vuv}), suggesting that terms with $z>3$ would be 
required to achieve a UV completion \cite{lif}.  In the theory with the 
extended gauge symmetry, the scalar mode is a gauge artifact, all physical 
modes aquire a $z=3$ dispersion relation at short distances from (\ref{vuv}), 
and no terms with $z>3$ are needed. 

The extended gauge symmetry of the theory with nonrelativistic general 
covariance has even more interesting consequences at long distances, because 
it improves the chances that the behavior of our theory can resemble general 
relativity in this observationally relevant regime.  We conclude this paper by 
previewing how our generally covariant theory compares to general 
relativity at long distances, focusing on the case of $3+1$ spacetime 
dimensions.   

Note first that even before we take the long-distance limit, the elimination 
of the scalar mode of the graviton is certainly a good sign for the possible 
matching against general relativity at long distances, and so is the fact that 
the coupling constant $\lambda$ in the kinetic term is now frozen by the 
symmetries of the generally covariant theory to take the relativistic value 
$\lambda=1$.  As a result, the number and the tensor structure of the 
gravitational wave polarizations is the same as in general relativity.  

In the infrared limit of our theory, the potential $\CV$ is dominated by the 
scalar curvature and the cosmological constant term, (\ref{vir}).  In 
this regime, the natural scaling is isotropic, with dynamical exponent $z=1$.  
The low-energy physics is best represented in rescaled coordinates $(x^0,x^i)$ 
and in terms of rescaled fields.  First, the new time coordinate 
\be
\label{lght}
x^0=\mu t
\ee
is defined by absorbing the effective speed of light $\mu$ into the definition 
of time.  Because $[\mu]=z-1$, this implies that $[x^0]=-1=[x^i]$, in accord 
with the $z=1$ scaling.  The rescaled fields are defined by
\be
\label{irf}
N^{\rm IR}_i=\frac{1}{\mu}N_i,\qquad A^{\rm IR}=\frac{1}{\mu^2}A.
\ee
This rescaling ensures (i) that $N^{\rm IR}_i$ carries the canonical dimension 
implied by $z=1$, and (ii) that the $U(1)$ gauge transformations are given by 
the standard relativistic formula
\be
\delta N^{\rm IR}_i=\p_i\alpha^{\rm IR},\qquad\delta A^{\rm IR}=\p_0
\alpha^{\rm IR},
\ee
with $\alpha^{\rm IR}=\alpha/\mu$.  In the rest of the paper, we will drop 
the ``IR'' superscripts, and refer to the rescaled fields (\ref{irf}) in the 
infrared simply as $N_i$ and $A$. 

The action of the infrared theory in the infrared variables is
\be
\label{sir}
S_{\rm IR}=\frac{1}{16\pi G_N}\int dx^0\,d^D\bx\,\sqrt{g}\left\{
\,N\left(K_{ij}K^{ij}-K^2+R-2\Lambda\right)-A(R-2\Omega)\vphantom{g^{ik}}
\right\}+\ldots,
\ee
where ``$\ldots$'' denotes corrections due to higher dimension operators, as 
well as the $\nu$-dependent terms in (\ref{fullact}) which are unimportant for 
our arguments below.  In (\ref{sir}), $K_{ij}$ refers to the extrinsic 
curvature tensor in the infrared coordinates, of canonical scaling dimension 
equal to one; and the Newton constant is given by
\be
G_N=\frac{\kappa^2}{32\pi\mu}.
\ee

In the remainder of this section, we comment on three issues:  The structure 
of compact-object solutions (which will be relevant for solar system tests), 
the issue of Lorentz symmetry, and the nature of cosmological solutions 
in the infrared regime of our theory as described by (\ref{sir}).   

\subsection{Static compact-object solutions}
\label{compob}

To prepare the ground for solar system tests, consider the infrared limit 
(\ref{sir}) and set the cosmological constant to zero.  Interestingly, as 
the Schwarzschild black hole turns out to be a solution of this infrared 
theory.  In terms of our fields, this solution will be represented by
\bea
\label{schwmet}
g_{ij}dx^idx^j&=&\left(1-\frac{2M}{r}\right)^{-1}dr^2+r^2d\Omega_2^2,\\
\label{schwrest}
A=1&-&\left(1-\frac{2M}{r}\right)^{1/2},\qquad N=1,\quad N_i=0,\quad \nu=0.
\eea
It is straightforward to see that this geometry satisfies the equations of 
motion of our theory for $\Omega=0$, which is the appropriate choice if we 
are interested in asymptotically flat solutions.  First, the equations of 
motion contain the condition $R=2\Omega$.  With $\Omega=0$, this equation is 
indeed satisfied by the spatial slices (\ref{schwmet}) of the relativistic 
Schwarzschild metric in the Schwarzschild coordinate system.  The $\nu$ and 
$N_i$ equations of motion are also satisfied, because the extrinsic curvature 
$K_{ij}$ vanishes for static backgrounds.  

Finally, to show that the $g_{ij}$ equation of motion are also satisfied, we 
use a simple but intriguing argument.  Since the same argument generalizes 
in a useful way to the case of nonzero $\Omega=\Lambda$, and also of arbitrary 
dimension, we present this more general case.  Start with static solutions 
with $K_{ij}=0$, and observe that the equations of motion for $g_{ij}$, $N$ 
and $A$ are identical to the equations that follow from the following reduced 
action, 
\be
\label{sred}
\int d^D\bx\,\sqrt{g}(N-A)(R-2\Omega).
\ee
Similarly, for static solutions with $K_{ij}=0$ of general relativity in 
$D+1$ dimensions, the corresponding equations of motion are those of the 
reduced Einstein-Hilbert action, 
\be
\int d^D\bx\,\sqrt{g}\,{\mathfrak N}\,(R-2\Lambda),
\ee
where $\mathfrak N$ is the general-relativistic lapse function.  Consequently, 
if we identify the $N$ and $A$ fields with the lapse function $\mathfrak N$ 
of general relativity, 
\be
\label{frna}
{\mathfrak N}=N-A,
\ee
we see that static solutions of general relativity are also solutions of our 
theory in the infrared limit.  In retrospect, this mapping also explains the 
form of $A$ in our representation of the Schwarzschild metric 
(\ref{schwrest}).  

Note that the relationship (\ref{frna}) between the general-relativistic 
lapse function $\mathfrak N$ and the $N$ and $A$ variables of our theory 
reproduces exactly what we would have expected from the geometric 
interpretation of $A$ as the subleading term in the expansion of the 
relativistic $g_{00}$ in powers of $1/c$ as obtained in (\ref{metrel}).  
Indeed, we start by expanding 
\be
\label{frm}
g_{00}\equiv-{\mathfrak N}^2=-(N-A)^2\approx -N^2+2NA+\ldots
\ee  
Recall now that $A$ in (\ref{frm}) is the infrared rescaled field 
(\ref{irf}), related to the microscopic gauge field by a rescaling factor 
$1/\mu^2$.  Using the fact that $\mu$ plays the role of the speed of light 
(as we have seen in (\ref{lght})), the two leading terms in (\ref{frm}) 
match exactly the leading two terms in the expansion (\ref{metrel}).  

These arguments prove that the Schwarzschild geometry in the Schwarzschild 
coordinates, with the indentification implied by (\ref{frna}), is a solution 
of the infrared limit of our theory, with $\Omega=\Lambda=0$.  However, in 
the parametrized 
post-Newtonian (PPN) formalism \cite{willb,will} which is typically used 
in gravitational phenomenology, the compact-object solution 
is usually represented in the isotropic coordinates.  In the case of 
general relativity, this is just a gauge choice, a fact which does not 
extend automatically to alternative approaches to gravity such as ours.  
Showing that the Schwarzschild geometry in the Schwarzschild coordinates is a 
solution of our theory does not imply that it will be a solution when 
represented in another coordinate system, because only those coordinate 
changes that belong to the gauge symmetry of our model will map a solution 
to a solution.  However, because the transformation from the Schwarzschild 
coordinates to the isotropic ones only changes the radial coordinate, 
\be
r=\rho\left(1+\frac{M}{2\rho}\right)^2,
\ee
while keeping $t,\theta,\phi$ intact, it is a foliation-preserving 
diffeomorphism, a symmetry of the theory.  Consequently, the Schwarzschild 
solution in the isotropic coordinates, represented by 
\bea
g_{ij}dx^idx^j&=&\left(1+\frac{M}{2\rho}\right)^{4}\left(d\rho^2
+\rho^2d\Omega_2^2\right),\\
A&=&\left(1+\frac{M}{2\rho}\right)^{-1}\frac{M}{\rho},\qquad N=1,\quad N_i=0,
\quad \nu=0,
\eea
is also a solution of the infrared limit of our theory.  Expanding this 
solution to the required order in the powers of $M/\rho$ strongly suggests 
that in the infrared regime, the $\beta$ and $\gamma$ parameters of the PPN 
formalism \cite{willb,will} will take the same values as in general 
relativity, $\beta=\gamma=1$.  This feature is favorable for the solar-system 
tests of the theory.  

\subsection{Lorentz symmetry}

Perhaps the leading challenge in any attempt to make theories of gravity with 
anisotropic scaling phenomenologically viable in $3+1$ dimensions is the  
issue of restoring Lorentz symmetry, at least at the intermediate energies and 
distances where it has been so well tested experimentally.  In particular, we 
need a mechanism ensuring that in the corresponding regime, all species of 
matter (including the gravitons) percieve the same lightcones and the same 
effective speed of light.  In the minimal theory with anisotropic scaling, 
this issue arises already for pure gravity:  At generic values of the 
couplings, the speeds of the tensor and scalar graviton polarizations are not 
related by any symmetry, and are generally different from each other already 
in the short-distance regime.  In contrast, our generally covariant theory 
has only the tensor graviton polarizations, all sharing the same speed at 
all energies; however, the issue reemerges when pure gravity is coupled to 
non-gravitational matter.  If the present theory is to be phenomenologically 
viable, its coupling to matter will have to be analyzed in detail.  This 
analysis is beyond the scope of the present paper; we only limit ourselves 
to one observation, which may be useful for the future analysis.  

In general relativity, Lorentz symmetry is a global symmetry associated with 
the isometries of the Minkowski spacetime.  In gravity with anisotropic 
scaling, we can adjust the couplings such that the flat spacetime geometry 
continues to be a solution.  The global symmetries of this solution will then 
depend on the precise model of gravity with anisotropic scaling.  

First consider the case of the minimal theory reviewed in 
Section~(\ref{secmin}), with the cosmological constant tuned to zero.  The 
flat spacetime (\ref{flatst}) is a solution, but it does not exhibit the full 
global Lorentz symmetry -- the Lorentz boosts, generated by 
\be
\label{lorentz}
\delta t=b_i x^i,\qquad \delta x^i=b_i t
\ee
(with $b_i$ a constant vector), are not foliation-preserving diffeomorphisms.  
In this theory, the nonrelativistic analogs of the Killing symmetries of the 
flat spacetime solution correspond to spacetime translations and space 
rotations -- the solution breaks all possible boost symmetries spontaneously, 
and defines a preferred rest frame.  

In contrast, in our generally covariant theory, we can interpret the Lorentz 
transformation (\ref{lorentz}) as a generator of a transformation belonging 
to the extended symmetry group $U(1)\ltimes\diff(M,\CF)$.  More precisely, 
the Lorentz transformation (\ref{lorentz}) should be interpreted as a 
composition of an infinitesimal foliation-preserving diffeomorphism and 
an infinitesimal $U(1)$ transformation.  Indeed, restoring the factors of $c$ 
shows that the variation of $t$ in (\ref{lorentz}) is suppressed by a factor 
of $1/c^2$ compared to the variation of $x^i$, and should therefore be 
interpreted as an infinitesimal $U(1)$ transformation with $\alpha=b_ix^i$, 
accompanied in (\ref{lorentz}) by the infinitesimal foliation-preserving 
diffeomophism 
\be
\delta t=0,\qquad\delta x^i=b_i t.
\ee
%
%%(
When interpreted in this way, the Lorentz transformation (\ref{lorentz}) is 
a symmetry of the flat spacetime geometry represented in our variables by 
$g_{ij}=\delta_{ij}$, $N=1$ and $A=0$.  This does not yet imply that all 
preferred-frame effects are absent in this background:  In particular, the 
Newton prepotential $\nu$ is not invariant under the Lorentz boosts, and 
defines a preferred frame for the flat spacetime, in which $\nu=0$.  The flat 
background is Lorentz invariant only to the extent that the effects of the 
Newton prepotential can be ignored.
%%)

\subsection{Cosmological solutions}

Moving beyond asymptotically flat spacetimes, it is natural to ask whether 
our theory has interesting cosmological solutions.  
One can start with a given spacetime geometry in general relativity, and 
investigate whether it satisfies the equations of motion of our theory.  
The answer to this question will again depend on the choice of spacetime 
foliation.  

For example, arguments identical to those used above for the Schwarzschild 
metric show that the static patch of the de~Sitter (or anti-de~Sitter) 
spacetime, represented in our variables by
\bea
g_{ij}dx^idx^j&=&\left(1-\frac{\Lambda r^2}{3}\right)^{-1}dr^2
+r^2d\Omega_2^2,\\
A=1&-&\left(1-\frac{\Lambda r^2}{3}\right)^{1/2},\qquad N=1,\quad 
N_i=0,\quad \nu=0,
\eea
is a solution of our theory if we set $\Omega=\Lambda$.  

It is encouraging to see that at least in the time-independent foliations, 
the de~Sitter and anti-de~Sitter spacetimes are solutions of our theory.  
In standard cosmological applications, however, the cosmological principle 
selects another natural foliation of spacetime, with homogeneous spatial 
slices and a time-dependent scale factor $a(t)$.  On the face of it, it may 
appear difficult to obtain cosmological solutions of our theory with maximally 
symmetric and time-dependent spatial slices:  The equation of motion for $A$ 
plays the role of a Gauss constraint, and implies $R=2\Omega$ in the vacuum.  
Assuming the standard FRW Ansatz
\be
\label{frw}
g_{ij}=a^2(t)\gamma_{ij},\qquad N=1,\qquad A=0,\qquad N_i=0,\qquad\nu=0
\ee
where $\gamma_{ij}$ is a time-independent maximally symmetric spatial metric, 
the scalar curvature of $g_{ij}$ has to be constant in time.  Consequently, if 
the scalar curvature of $\gamma_{ij}$ is nonzero, the cosmological scale factor 
must be independent of time.

Of course, if our spatial slices are flat, the Gauss constraint no longer 
restricts the time dependence of the cosmological scale factor.  This requires 
$\Omega=0$.  The rest of the equations of motion will be satisfied by the 
de~Sitter spacetime in the inflationary coordinates, which in the FRW Ansatz 
(\ref{frw}) corresponds to
\be
a(t)=e^{Ht},\qquad \gamma_{ij}=\delta_{ij}.
\ee
The reason for this is again simple but illuminating:  With $\Omega=0$, the 
$\nu$ equation of motion is satisfied when the metric is flat.  With $\nu$ 
and $A$ both zero, the remaining equations are implied by Einstein's 
equations if we simply identify the relativistic lapse function with our 
$N(t)$, the relativistic cosmological constant with our $\Lambda$, and $H$ 
with the Hubble constant.  

Thus, we see that the same de~Sitter spacetime in two different foliations 
is a solution of the infrared theory for two different choices of the coupling 
constant, one with $\Omega=\Lambda$ and the other with $\Omega=0$ and nonzero 
$\Lambda$.  Mapping out the general behavior of cosmological solutions as the 
coupling constants $\Omega$ and $\Lambda$ are independently varied is one of 
questions left for future work.  

In addition, there are at least two ways out of the potential difficulty with 
solving the Gauss constraint for cosmologically evolving spacetimes with 
maximally symmetric spatial slices of nonzero curvature.  First, the equations 
of motion will change in the presence of matter.  In the full system of 
equations for gravity and matter, the Gauss constraint is expected to be 
modified by a matter source, whose time dependence can then drive the time 
dependence of the scale factor in the spatial metric.  The second possibility 
is related to the gauge freedom we have in describing cosmological solutions 
in general relativity:  Instead of the standard FRW ansatz which leads to 
(\ref{frw}), one can choose coordinates in which the spatial metric is not 
only maximally symmetric but also constant in time.  When we express the FRW 
geometry in such coordinates, the time-dependent scale factor appears in the 
$dt^2$ term in the metric, and non-zero components of the shift vector $N_i$ 
are typically generated.  In general relativity, this coordinate 
representation of FRW cosmologies is a legitimate albeit slightly 
unconventional gauge choice.  In our theory, this parametrization of FRW 
universes has the advantage of being compatible with the vacuum Gauss 
constraint $R=2\Omega$.  

\acknowledgments

\hglue.4in 
We wish to express our thanks to Niayesh Afshordi, Jan Ambj\o rn, Dario 
Benedetti, Diego Blas, Robert Brandenberger, Gia Dvali, Marc Henneaux, Elias 
Kiritsis, Renate Loll, Alex Maloney, Shinji Mukohyama, Yu Nakayama, Oriol 
Pujol\`{a}s, Sergey Sibiryakov, Arkady Vainshtein, and the participants of the 
Perimeter Institute workshop on {\it Gravity at a Lifshitz Point\/} (November 
2009) for useful discussions.  P.H. is grateful to the Arnold Sommerfeld 
Center for Theoretical Physics, Ludwig-Maximilians-Universit\"at, M\"unchen, 
and the PH-TH Division, CERN, Gen\`{e}ve, for their hospitality during some 
of the final stages of this work.  The results of this work were presented at 
the {\it GR 19 Conference\/} in Mexico City in July 2010; P.H. wishes to 
thank the organizers for their invitation and hospitality.  This work has been 
supported by NSF Grant PHY-0855653, DOE Grant DE-AC02-05CH11231, and by the 
Berkeley Center for Theoretical Physics.  

\bibliographystyle{JHEP}
\bibliography{gen2}
\end{document}